\documentclass[11pt]{article}
\usepackage[dvips]{graphicx}
\begin{document}
\begin{center}
{\LARGE\bf Cosmology in Nonlinear Born-Infeld Scalar Field Theory
With Negative Potentials}
 \vskip 0.15 in
 $^\dag$Wei Fang$^1$, $^\ddag$H.Q.Lu$^1$, Z.G.Huang$^2$\\
$^1$Department~of~Physics,~Shanghai~University,\\
~Shanghai,~200444,~P.R.China\\
$^2$Department of Mathematics and Science,
\\Huaihai Institute of Technology, Lianyungang, 222005, P.R.China
\footnotetext{$\dag$ xiaoweifang$_-$ren@shu.edu.cn}
\footnotetext{$\ddag$ Alberthq$_-$lu@staff.shu.edu.cn}

 \vskip 0.5 in
\centerline{\bf Abstract} \vskip 0.2 in
\begin{minipage}{5.5in} { \hspace*{15pt}\small \par The cosmological evolution in Nonlinear Born-Infeld( hereafter NLBI)
scalar field theory with negative potentials was investigated. The
cosmological solutions in some important evolutive epoches were
obtained. The different evolutional behaviors between NLBI and
linear(canonical) scalar field theory have been presented. A
notable characteristic is that NLBI scalar field behaves as
ordinary matter nearly the singularity while the linear scalar
field behaves as "stiff" matter. We find that in order to
accommodate current observational accelerating expanding universe
the value of potential parameters $|m|$ and $|V_0|$ must have an
{\it upper bound}. We compare different cosmological evolutions
for different potential parameters $m, V_0$.

 }
\end{minipage}
\end{center}
\newpage
\section{Introduction}
 \hspace*{15pt}\par The role of rolling homogeneous scalar field has been
 widely discussed in the various epoch for a variety of
 purposes[1]. Recently, with the surprising discovery of an
 accelerating expansive and spatially flat universe, the scalar
 field has gained another newly discussion as a candidate for dark
 energy. It can drive current accelerating expansion while its
 energy density can fill in the universe as "missing matter
 density". The most popular models with scalar field may be the
 linear scalar field model( a canonical scalar field described by the lagrangian
 $L=\frac{1}{2}\dot{\phi}^2-V(\phi)$)[2-6], the K-essence model( a scalar field with a non-canonical kinetic energy
 terms)[7-22] and the "phantom" model(a scalar field with the negative kinetic energy
 terms)[23-52]. The potentials in these models are chosen non-negative
 to avoid negative potential energy density. It is shown that the
 expanding universe with non-negative potentials have a common
 property that they will expand for ever, though the evolutional
 behavior of future universe has significant differences
 corresponding to different potentials. However, research shows that
 negative potentials can also lead to a viable cosmology[53-56]. Moreover,
 the universes with negative potentials are entirely different with the universes with non-negative potentials. They can
 trigger our flat universe from expansion($H>0$) to contraction
 ($H<0$), which will never occur in standard FRW model( previous
 oscillatory model only appears in a close universe in standard
 FRW model). Hence negative potentials are used to propose the "cyclic
 universe" model. In this cyclic scenario, when the scalar field
 rolls to a minimum of its effective potential with $V(\phi)<0$,
 the universe will stop expanding and contract to a singularity eventually. Additionally, negative
 potentials also appeared in supergravity theory and in brane cosmology. It is
 theoretically important to continue investigating the
 cosmological features in other models where the effective
 potential $V(\phi)$ may become negative for some values of the
 field $\phi$.
 \par Nonlinear Born-Infeld scalar field theory is firstly proposed by W.Heisenberg in order to describe the process
 of meson multiple production connected with strong field
 regime[57-59] and then is discussed in cosmology[60-64]. It shows that
 the lagrangian density of this NLBI scalar field posses some
 interesting characteristics[65-66]. In Ref[65], the author showed that
 a singular horizon exists for a large class of solution in which the scalar
 field is finite. Naked
 singularities with everywhere well-behaved scalar field in
 another class of solution have also been found in Ref[65]. Lately the
 quantum cosmology with the NLBI scalar field has been
 considered[67]. In the extreme limits of small and large
 cosmological scale factor the wave function of the universe was
 found by applying the methods developed by Vilenkin, Hartle and
 Hawking. The result has suggested a non-zero positive
 cosmological constant with largest probability, which is
 consistent with current observational data. The classical
 wormhole solution and wormhole wavefunction with the NLBI scalar
 filed has been obtained in Ref[68]. The phantom
 cosmology based on NLBI scalar field with a special potential had been considered in Ref[69-70].
 The results show that the universe will evolve to a de-sitter
 like attractor regime in the future and the phantom NLBI scalar
 field can survive till today without interfering with the
 nucleosynthesis of the standard model. Very recently, with the
 analysis to Gold supernova data, we show that maybe the NLBI
 scalar field model is superior to conventional quintessence
 model[71]. Furthermore it is showed that in another analogous NLBI theories with the
 lagrangian $p(\phi,X)=\alpha^2(\sqrt{(1+\frac{2X}{\alpha^2}}-1)-\frac{1}{2}m^2\phi^2$(where $X=\frac{1}{2}\dot\phi^2$) the contribution of the gravitational
waves to the CMB fluctuations can be substantially larger than that
naively expected in simple inflationary models, which make the
prospects for future detection much more promising[72-73]. It is
also showed that with the same lagrangian, one can send information
from inside a black hole[74]. In Ref[75-76], authors consider a
non-Abelian Einstein-Born-Infeld-dilaton theory, where they concern
a non-abelian vector field which couples to the dilaton and then
describe a dark energy mechanism in a cosmological framework.
 \par The key idea of NLBI scalar field theory is that the
 conventional quintessence scalar field can not describe the
 reality correctly in the case of strong field. The lagrangian of
 conventional quintessence model(here we also call it linear scalar
 field):\begin{equation}L=\frac{1}{2}{\dot\phi}^2-V(\phi)\end{equation}
 should be substituted by the lagrangian of NLBI scalar field
  \begin{equation}L_{NLBI}=\frac{1}{\eta}[1-\sqrt{1-\eta{\dot\phi}^2}]-V(\phi)\end{equation}
  which can recover to conventional case when $\dot\phi\rightarrow
  0$. In fact, the lagrangian of NLBI scalar field(Eq.2) implies
  that there exists a maximum constant value
  $\frac{1}{\sqrt{\eta}}$ for field velocity $\dot\phi$, which is
  very analogous to the universal constant velocity $c$. It means that $\dot\phi$
never reaches infinity while in linear scalar field model there
are no such constraint.
\par In this paper, we combine the two
ideas(negative potentials and NLBI scalar field theory) and
consider the cosmology based on the NLBI scalar field with
negative potentials. We think it may be very interesting and
meaningful to see what will happen in this case. The paper is
organized as follows:
   In section 2 we will describe theoretical model in NLBI scalar field theory and consider several basic regimes which are
   possible to happen in NLBI scalar field: the potential energy dominated regime, the kinetic energy dominated
   regime and the transient regime that the universe switches from expansion to
contraction. In section 3, we investigate the different
cosmological evolution in different cases and plot corresponding
evolutive behaviors in detail. For the potential
$V(\phi)=\frac{1}{2}m^2\phi^2+V_0(V_0<0)$ We consider the universe
evolution with different slope $m$ and different potential well
$V_0$. the cases that $V_0>0$ and $V_0=0$ are also presented to
compare with the case $V_0<0$, moreover we compare the different
evolution between NLBI scalar field and linear scalar field. in
section 4 we mention the cyclic model and consensus model.
Conclusion and summary is also presented in section 4.
\section{Theoretical Model in NLBI Scalar Field theory}
We consider the behavior of the NLBI scalar field in the Friedmann
universe with the spatially flat FRW metric
$ds^2=dt^2-a^2(t)(dx^2+dy^2+dz^2)$. The energy density and
pressure density for NLBI scalar field
are:\begin{equation}P_{NLBI}=\frac{1}{\eta}[1-\sqrt{1-\eta{\dot\phi}^2}]-V(\phi)\end{equation}
\begin{equation}\rho_{NLBI}=\frac{1}{\eta\sqrt{1-\eta{\dot\phi}^2}}-\frac{1}{\eta}+V(\phi)\end{equation}
Corresponding Friedmann equation is
\begin{equation}H^2=\frac{1}{3{M_p}^2}(\rho_{NLBI}+\rho_{\alpha})=\frac{1}{3{M_p}^2}[\frac{1}{\eta\sqrt{1-\eta{\dot\phi}^2}}-\frac{1}{\eta}+V(\phi)+\rho_{\alpha}]\end{equation}
$\rho_{\alpha}$ is the  energy density of a matter with baryotropic
equation of state $p_{\alpha}=\alpha \rho_{\alpha}$, where $\alpha$
is a constant. For nonrelativistic matter $\alpha=0$, for radiation
$\alpha=\frac{1}{3}$. The evolution equations with Hubble parameter
$H$ are:
\begin{equation}\dot H=-\frac{1}{2{M_p}^2}(\rho_{NLBI}+\rho_{\alpha}+p_{NLBI}+p_{\alpha})
=-\frac{1}{2{M_p}^2}[\frac{\dot\phi^2}{\sqrt{1-\eta{\dot\phi}^2}}+(1+\alpha)\rho_{\alpha}]\end{equation}
\begin{equation}\dot{\rho}_{\alpha}=-3H(\rho_{\alpha}+p_{\alpha})\end{equation}
\begin{equation}\ddot\phi+(1-\eta{\dot\phi}^2)[3H\dot\phi+\frac{dV(\phi)}{d\phi}(1-\eta{\dot\phi}^2)^{\frac{1}{2}}]=0\end{equation}
When $\dot{\phi}\rightarrow 0$, ignoring the higher-order term of
$\dot\phi$ Eq.(8) will recover to quintessence model:
\begin{equation}\ddot\phi+3H\dot\phi+\frac{dV(\phi)}{d\phi}=0\end{equation}
where potential $V(\phi)=\frac{1}{2}m^2 \phi^2+V_0(V_0<0)$. Eq.(8)
also tells us that when $\dot\phi$ increase to its maximum
$\frac{1}{\sqrt\eta}$, $\ddot\phi$ will decrease to zero and this
prevents $\dot\phi$ from increasing continuously.
\par We will use a system of units in which $M_p=(8\pi G)^{-1/2}=\eta=1$ for convenience.
\par It is very difficult to
obtain the general exact solution for the Eqs.(6-8). However we are
able to obtain the solutions in some special regimes: \par
\textbf{A.The potential regime: Energy density dominated by
$V(\phi)$}\par In this case $\dot\phi^2/2, \rho_{\alpha}<<V(\phi)$
and $|\ddot\phi|<<|3H\dot\phi|$. We will find that in this case the
result for NLBI scalar field is similar to the one shown in Ref[77].
It corresponds to the vacuum-like equation of state:
\begin{equation}p=-V(\phi)=-\rho\end{equation}
The equations for $a$ and $\phi$ in this regime have the following
form:\begin{equation}H^2=(\frac{\dot
a}{a})^2=\frac{m^2\phi^2}{6}+\frac{V_0}{3}\end{equation}
\begin{equation}3H\dot\phi+m^2\phi(1-{\dot\phi}^2)^{\frac{1}{2}}=0\end{equation}
When $m^2\phi^2>>|V_0|$, we can obtain the solutions for $\phi$
and $a$:
\begin{equation}\phi(t)=\phi_0-\sqrt{\frac{2}{2m^2+3}}mt\end{equation}
\begin{equation}a(t)=a_0 e^{\sqrt{\frac{2m^2+3}{48}}({\phi_0}^2-\phi^2(t))}\end{equation}
Where $\phi_0$ and $a_0$ are different integral constant. These
solution are very anologous to the results in Ref[77-79]. It has
been argued that these solutions describe inflationary universe.
However if we consider positive potential
$V(\phi)=\frac{1}{2}m^2\phi^2+V_0(V_0>0)$ and
$V_0>>\frac{m^2\phi^2}{2}$. From Eq.(12), we can get
\begin{equation}\frac{d\phi^2}{dt}=-\frac{2m^2\phi^2}{\sqrt{m^4\phi^2+3V_0}}\end{equation}
Integrating Eq.(15), we get the solutions for $\phi$ and $a$:
\begin{equation}
t+\frac
{\sqrt{m^4\phi^2+3V_0}}{m^2}-\frac{\sqrt{3V_0}arctanh\sqrt{\frac{m^4\phi^2+3V_0}{3V_0}}}{m^2}+C_1=0
\end{equation}
\begin{equation}a(t)=a_0e^{\sqrt{\frac{V_0}{3}}t}\end{equation}
where $C_1$ is the integral constant. For linear scalar field, the
corresponding equation is
\begin{equation}\frac{d\phi^2}{dt}=-\frac{2m^2\phi^2}{\sqrt{3V_0}}\end{equation}
and the solutions are
\begin{equation}\phi=\phi_0e^{-m^2t/\sqrt{3V_0}},  a(t)=a_0e^{\sqrt{\frac{V_0}{3}}t}\end{equation}
Eqs.(16,17) show the evolution of universe when $\phi$ is in the
bottom of potential while Eqs.(13,14) describe the universe when
$\phi$ is far from the bottom. This two regime are both the
accelerating expansion, so it can be considered as a simple
version to describe an eternally self-reproducing inflationary
universe, as well as the present stage of accelerating expansion.
Eq.(19) shows that $\phi$ will roll down the potential and settle
on the bottom of the potential permanently to mimic the de-sitter
accelerating expansion. From Eqs.(15,18), we can also know, due to
the nonlinear effect, the attenuation of NLBI scalar field is
slower than linear scalar field. However if $V_0<0$, the later
time accelerating expansion will never occur. The evolution of
universe will be completely different with the case $V_0>0$
\par \textbf{B.
The kinetic Regime:Energy Density Dominated by Kinetic Energy}
\par This regime is very important because in this case the
nonlinear effects are distinct. This regime corresponds to strong
field $\dot\phi$. When energy density dominated by kinetic energy,
we can neglect $V(\phi)$ and $\rho_{\alpha}$. Then the Eqs.(5,8)
become:
\begin{equation}H^2=\frac{1}{3}(\frac{1}{\sqrt{1-{\dot\phi}^2}}-1)\end{equation}
\begin{equation}\ddot\phi+3H\dot\phi(1-{\dot\phi}^2)=0\end{equation}
we have the solution
\begin{equation}\dot\phi=\pm\sqrt{\frac{1}{1+a^6}}\end{equation}
Since in the kinetic energy dominated epoch, $\dot\phi$ is close
to $1$, from Eq.(22) we know that the scalar factor $a(t)$ will be
very small. Eq.(22) describes an expanding universe from a
singularity or a contracting universe towards a singularity. Form
Eqs.(20,22), we can obtain
\begin{equation}\rho\sim
H^2=\frac{\sqrt{1+a^6}}{a^3}-1 \sim
\frac{1}{a^3}+\frac{1}{2}a^3-1\sim \frac{1}{a^3}\end{equation} The
solutions can be written as follows for small scale factor and
strong field:\begin{equation}a(t)\sim t^{2/3}\end{equation}
\begin{equation}\dot\phi=\pm\sqrt{\frac{1}{1+a^6}}=\pm\sqrt{\frac{1}{1+t^4}}\approx\pm(1-\frac{1}{2}t^4)\end{equation}
\begin{equation}\phi=\phi_0\pm(t-t_0)\mp\frac{1}{10}(t-t_0)^5 \end{equation}
We can represent the kinetic energy term and the potential energy
term as:$E_k\sim H^2\sim\frac{1}{t^2}$ and
$E_p=V(\phi)\sim\phi^2$. Therefore we can conclude that: Firstly,
if the solution describes an expanding universe from a
singularity, $E_k$ drops down rapidly while $E_p$ changes slowly.
Therefore the regime with energy density dominated by potential
$V(\phi)$ will appear when the evolution is far from the
singularity. Secondly, if the solution describes a contracting
universe towards a singularity, $E_p$ grows slowly while $E_k$
diverges as $t^{-2}$, ($t$ is the time remaining before the big
crunch singularity), and in this case $E_k$ will dominate the
universe. Therefore we can conclude that the kinetic energy $E_k$
will always dominate the universe in the vicinity of the
singularity.
\par Here we should point out the different results between NLBI
scalar field and linear scalar field. The solution for $a(t)$ and
$\phi(t)$ of linear scalar field are given in[77]: $a(t)\sim
t^{1/3}$, $\phi=\phi_0 \pm \sqrt{\frac{2}{3}}ln\frac{t_0}{t}$,
$\frac{\dot\phi^2}{2}=\frac{1}{3t^2}$, which are different to the
solutions Eqs.(24-26). The NLBI scalar field behaves like
nonrelativistic matter ($a(t)\sim t^{2/3},\rho\sim \frac{1}{a^3}$)
near the singularity while the linear scalar field behaves like
"stiff" matter($a(t)\sim t^{1/3},\rho\sim \frac{1}{a^6}$). This may
lead to some interesting cosmological implies. However we should
point out that if the evolution happens in the presence of other
fields or other source of matter whose density energy behaves as
$\frac{1}{a^n}(n>3$), then our NLBI scalar field will not dominate
the universe near the singularity. In this case the evolution
presented here will be modified by the influence of other fields
near the singularity.
\par \textbf{C.The transient regime: Switch from expansion to contraction}
After analyzing two special regimes, we now pay attention to
another important regime: the transient regime that the universe
begin to contraction from a expanding phase. Before numerically
studying this process, we try to get some solutions by some simple
approximation. Since we study the very vicinity where Hubble
parameter vanishes($H\sim0$), we can neglect the term
$3H\dot\phi(1-\dot\phi^2)$ and rewrite Eq.(8) as:
\begin{equation}\ddot\phi+m^2\phi(1-{\dot\phi}^2)^{3/2}=0\end{equation}
Furthermore, the value of $\dot\phi$ will be also very small in
this case. Taking the first order approximation of
$(1-{\dot\phi}^2)^{3/2}$ in Eq.(27), we get:
\begin{equation}\ddot\phi+m^2\phi(1-\frac{3}{2}{\dot\phi}^2)=0\end{equation}
Integrating Eq.(28), we obtain:
\begin{equation}\dot\phi^2=\frac{2-e^{\frac{3}{2}m^2\phi^2}}{3}\end{equation}
In Ref[77], it is argued that only if $|V_0|\sim 10^{-120}$ the time
that universe begins to collapse can be greater than the age of
present universe. In the transient regime, $\frac{3}{2}m^2\phi^2
\sim |V_0|\sim10^{-120}$, so the term $ e^{\frac{3}{2}m^2\phi^2}$
can be taken as $1+\frac{3}{2}m^2\phi^2$.  Eq.(29) can be written as
\begin{equation}3\dot\phi^2+\frac{3}{2}m^2\phi^2=1\end{equation}
The solution of Eq.(30) is
\begin{equation}\phi\propto\cos\frac{\sqrt2}{2}mt\end{equation}
\par Up to now, by some reasonable approximation, we know that in
the vicinity where the universe evolves to contraction from
expansion, the field $\phi$ will experience a simple oscillatory
motion $\phi\propto \cos\frac{\sqrt2}{2}mt$(Fig.2). With respect to
the equation in Ref[77], where $\phi\propto \cos mt$, it is the
nonlinear effect that makes the different evolution.
\par Next we will investigate what happens during the transient regime (where the sign of $\dot a$ changes). We also try to obtain the
analytical solution by some reasonable approximations
 \par First
of all, we represent $V(\phi)=\frac{1}{2}m^2\phi^2-|V_0|$ in the
form $V(\phi)=\frac{1}{2}m^2(\phi^2-{\phi_0}^2)$ for convenience.
we assume that the field $\phi$ begins the oscillation at $t=0$,
moving with zero initial velocity from a point
$\phi_1\approx\phi_0$. The initial energy density of the field is
$\triangle
V=V(\phi_1)=\frac{1}{2}m^2({\phi_1}^2-{\phi_0}^2)<<|V(0)|$. We
will evaluate the turning point moment $t_c$ when $H\sim0$(i.e,
$\dot a\sim0)$.
\par Using the same method in Ref[77], we will consider the series
expansion of the Hubble parameter around the beginning of this
process
\begin{equation}
H(t)=H_1+H^{(1)}_1t+\frac{1}{2!}H^{(2)}_1t^2+\frac{1}{3!}H^{(3)}_1t^3+\cdots\end{equation}
where $H_1$ and $H^{(n)}_1$ are taken the value of $H$ and
$H^{(n)}=\frac{d^nH}{dt^n}$ at $t=0$. From Eq.(6) we have $\dot
H=-\frac{\dot\phi^2}{2\sqrt{1-\phi^2}}$(here we ignore the
presence of baryotropic matter $\rho_\alpha$), then we find that
$H^{(1)}_1=H^{(2)}_1=0$ for vanishing initial velocity
$\dot\phi=0$ at $t=0$. The first non-vanishing coefficient $
{H_1}^{(3)}=-\ddot\phi^2(1-\dot\phi^2)^{-\frac{1}{2}}=-V'^2(1-\dot\phi^2)^{5/2}=-V'^2=m^4\phi^2(0)$.
Including the terms up to $t^3$ in Eq.(32), we get:
\begin{equation}t_c=(\frac{12V(\phi_1)}{V'(\phi_1)^4})^{1/6}=m^{-1}(\frac{12\triangle V}{m^2{\phi_0}^4})^{1/6}\end{equation}
This means that the Hubble parameter vanishes at the time $t_c$,
where
\begin{equation}\phi_c=\phi_0-(\frac{3\triangle V}{16m^2\phi_0})^{1/3}\end{equation}
Here the result of $t_c$(Eq.(33)) is the same as Ref[77], this is
because we set the initial velocity $\dot\phi(0)=0$. When $\triangle
V<m^2{\phi_0}^4$, this results imply that the turn occurs in the
vicinity of the point $\phi_0$ where the potential becomes negative.
In the short time when universe begins to contract from expansion,
we can also study the subsequent evolution of $\phi(t)$ and $a(t)$.
We therefore take $a=1$ during this time, and
$\phi(t)=\phi_1\cos\frac{\sqrt2}{2}mt$. The potential can be
expressed as:
\begin{equation}V(\phi)=\triangle
V-\frac{m^2{\phi_1}^2}{2}\sin^2\frac{\sqrt2}{2}mt\end{equation}
The acceleration of the universe is given by
\begin{equation}\ddot a\simeq\frac{\triangle V}{3}-\frac{1}{3}m^2{\phi_1}^2\sin^2\frac{\sqrt2}{2}mt
\\-\frac{1}{16}m^4{\phi_1}^4\sin^4\frac{\sqrt2}{2}mt\end{equation}
The initial value of $\dot a$ equals $\dot a=\pm a\sqrt{\triangle
V/3}=\pm\sqrt{\triangle V/3}$, this yields
\begin{equation}\begin{array}{lll}\dot a=\pm\sqrt{\frac{\triangle V}{3}}+\frac{1}{3}\triangle Vt-\frac{1}{6}m^2{\phi_1}^2t+\frac{\sqrt2}{12}m{\phi_1}^2\sin(\sqrt2mt)
   \\-\frac{3}{128}m^4{\phi_1}^4t+\frac{\sqrt2}{64}m^3{\phi_1}^4\sin(\sqrt2mt)-\frac{\sqrt2}{512}m^3{\phi_1}^4\sin(2\sqrt2mt)\end{array}\end{equation}
 Where the sign $"+"$ denotes an expanding universe at the beginning of the oscillation.
 In this case the universe will stop its expansion at $\phi=\phi_c$ and then collapse to singularity.
 $"-"$ denotes a collapsing universe at the beginning of the oscillation. In this time the universe will continue collapsing to singularity(see Fig.1).
 We numerically plot  the
evolution of the scale factor $a$ and scalar field $\phi$(Fig.1
and Fig.2) at the time when universe switches from expansion to
contraction.
 \vskip 0.3 in
\begin{center}
\begin{minipage}{0.45\textwidth}
\includegraphics[scale=0.32,origin=c,angle=270]{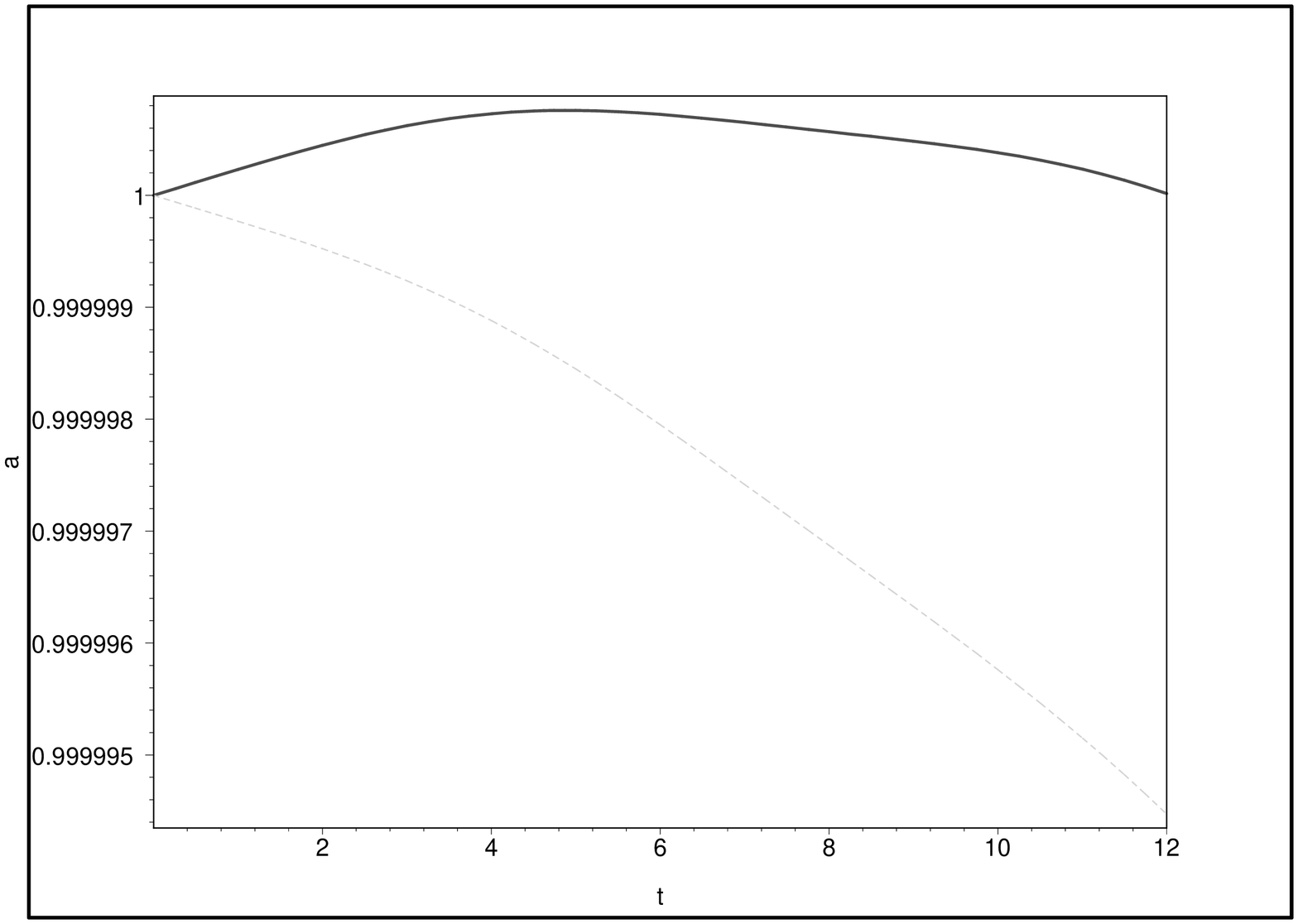}
{\small  ~Fig1.The evolution of scalar factor when $H\sim0$. Solid
line describes an expanding universe at the beginning of the
oscillation. Clearly it stops its expansion and collapses to
singularity. dash line describes a collapsing universe, it
continues collapsing to singularity.}
\end{minipage}
\hfill
\begin{minipage}{0.50\textwidth}
\includegraphics[scale=0.32,origin=c,angle=270]{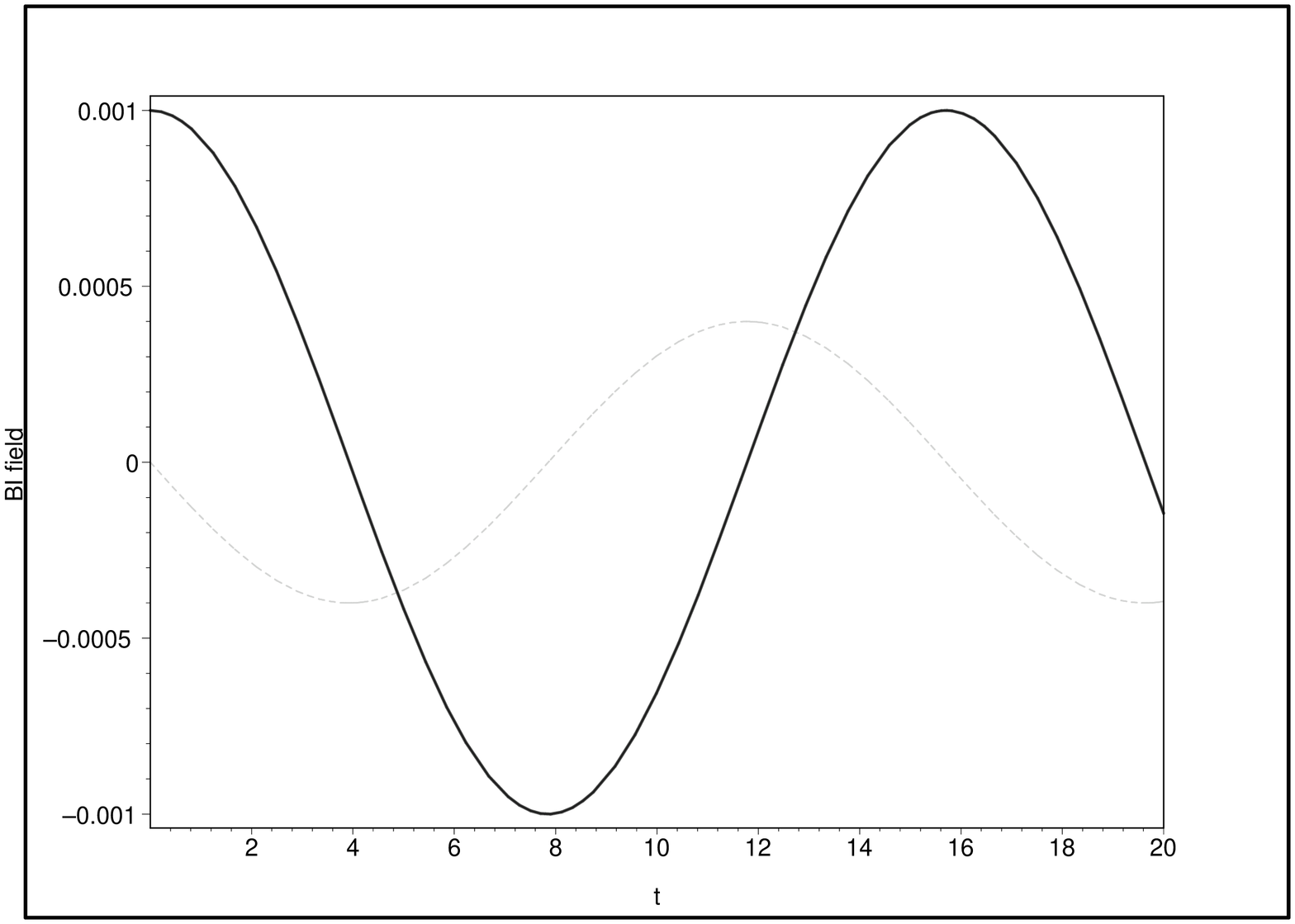}
{\small ~Fig2.The evolution of $\phi$ and $\dot\phi$ when
$H\sim0$. Solid line is the evolution of $\phi$, dash line is the
evolution of $\dot\phi$. From the figure we know that the
approximate solution Eq.(31) is reliable. We set $\phi_0=0.001$,
the initial value of $\phi(0)=0.001000001$, $m=0.4$,
$\dot\phi(0)=0$, $a(0)=1$.}
\end{minipage}
\end{center}

\section{The Cosmological Evolution in NLBI Scalar Field and Linear Scalar
Field Theory} \par In this section, we will investigate the
cosmological behaviors using numerical approach and plot the
results in details. For the potential
$\frac{1}{2}m^2\phi^2+V_0(V_0<0)$, we will consider the different
cosmological evolutions with different potential wells(i.e,
different negative $V_0$ value) and different potential slope
(i.e, different $m$ value). We will also plot the different
behaviors between NLBI scalar field and linear scalar field
theory. The different evolutions in case of $V_0>0, V_0=0, V_0<0$
are also studied. In fact, the evolutions of these three cases
describe the common feature of three class of potentials: positive
potentials, non-negative potentials and negative potentials.
\par \textbf{Case 1. Same slope $m$ but different potential well $V_0$}
\begin{center}\vspace{0.5cm}
\begin{minipage}{0.50\textwidth}
\includegraphics[scale=0.32,origin=c,angle=270]{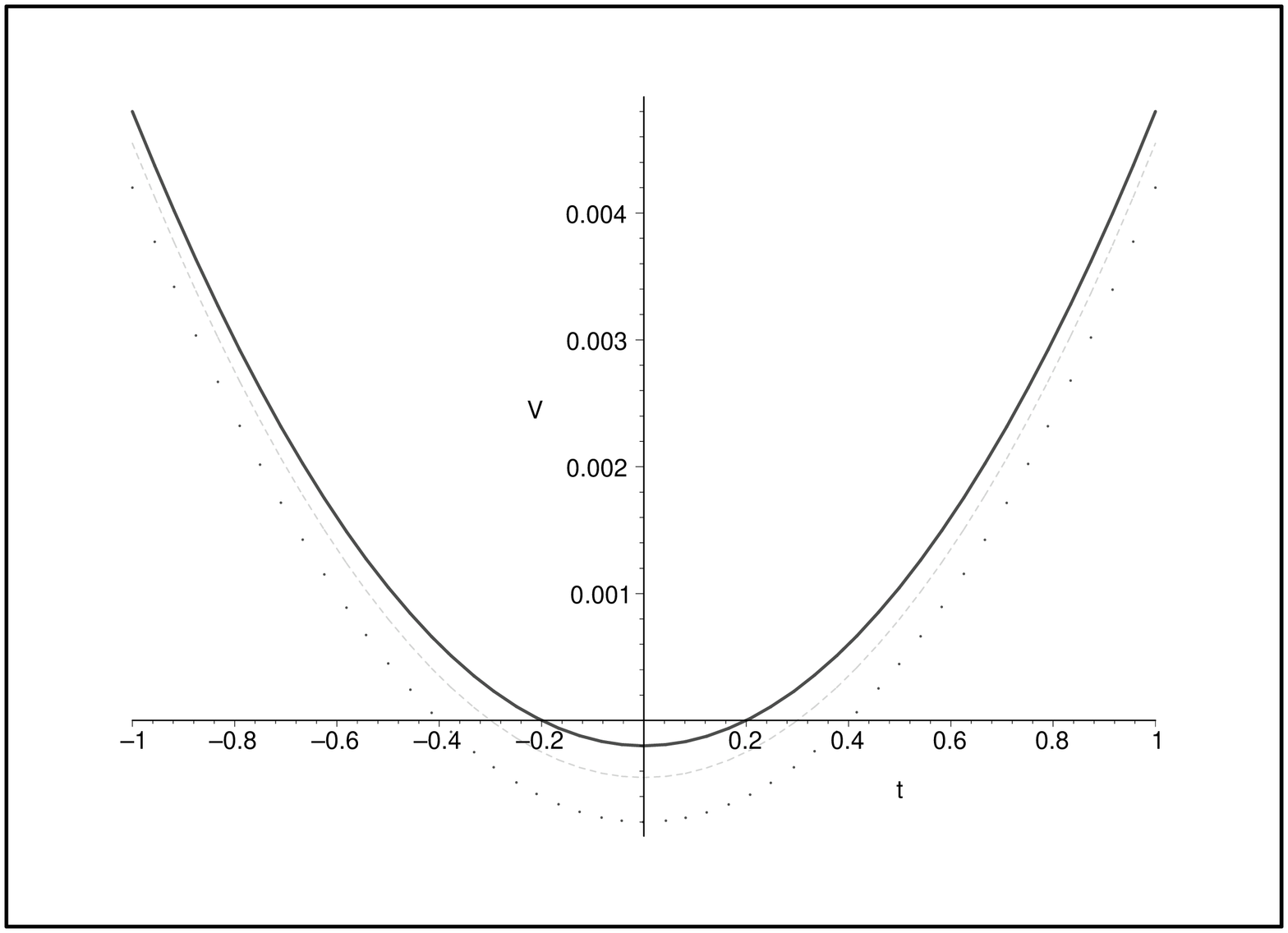}
{\small~Fig3. The three potentials with same slope $m(=0.1)$ but
different potential well $V_0$. $V _0=-2.0\times10^{-4}$ for solid
line ,$V _0=-4.5\times10^{-4}$ for dash line and  $V
_0=-8.0\times10^{-4}$ for dot line.\\}
\end{minipage}
\hfill
\begin{minipage}{0.45\textwidth}
\includegraphics[scale=0.32,origin=c,angle=270]{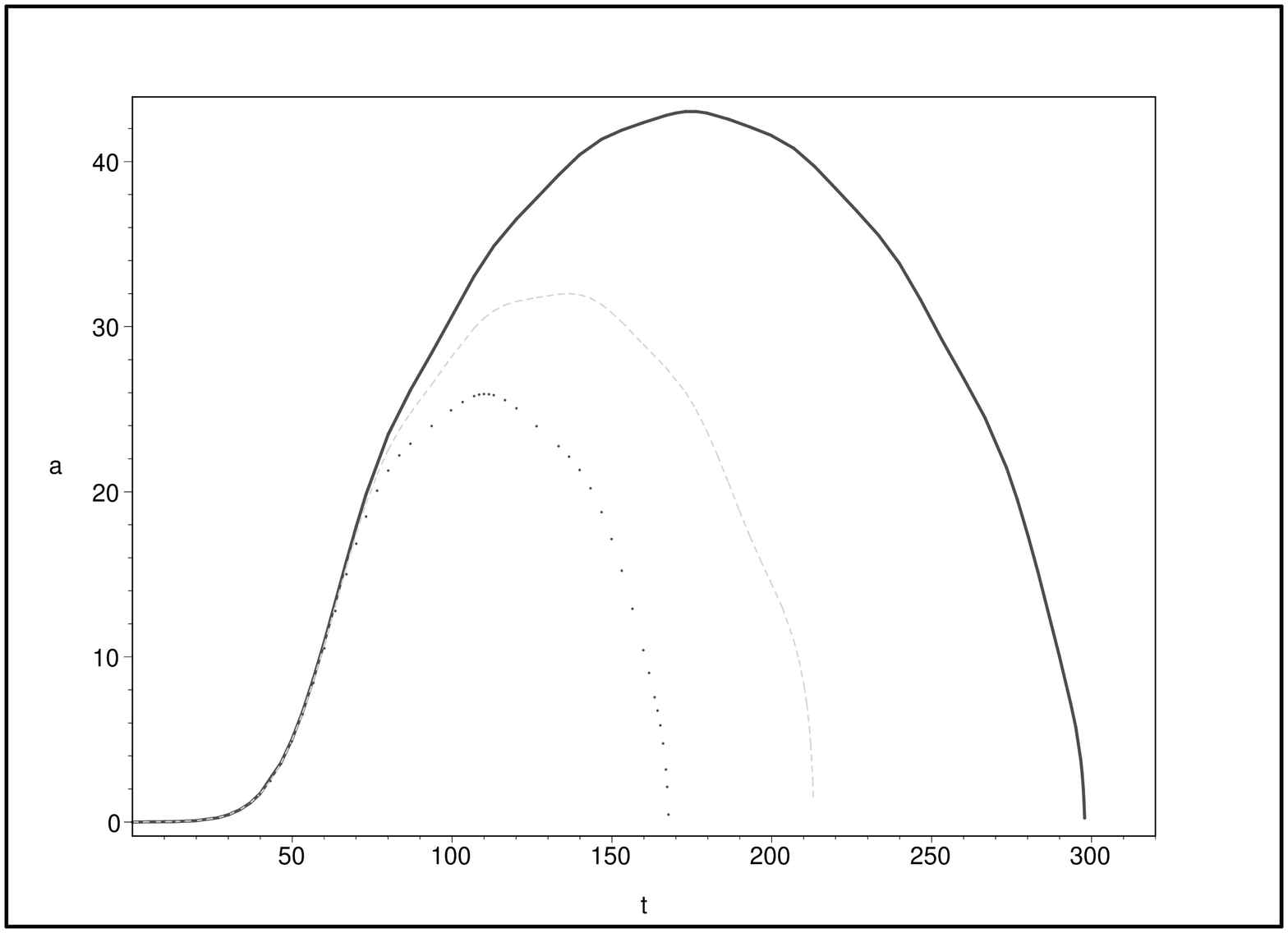}
{\small~Fig4. The evolution of scale factor $a(t)$ with the three
potentials. The value of $V_0$ is the same as Fig.3 and the initial condition is $\phi(0)=6, \dot\phi(0)=0$\\
}
\end{minipage}
\end{center}
\par From Fig.4, it shows that the universe can undergo an accelerating
expansion,  then a decelerating expansion and ultimately contract
to singularity. Though the potential well are different, the
universe have nearly same evolution at the beginning. It also
shows that the deeper the potential well is, the shorter the age
of universe is. Since we now live in a expanding universe where
the large scale structure had formed, there must exist a upper
bound for the value of potential well $|V_0|$. At the time when
universe begins to contract from an expansion, the scalar field
$\phi$ begins to oscillate and ultimately moves to
$-\infty$(Fig.5). The velocity of field $|\dot\phi|$ will reach
its maximum value $1$ finally.
\begin{center}\vspace{0.5cm}
\begin{minipage}{0.47\textwidth}
\includegraphics[scale=0.32,origin=c,angle=270]{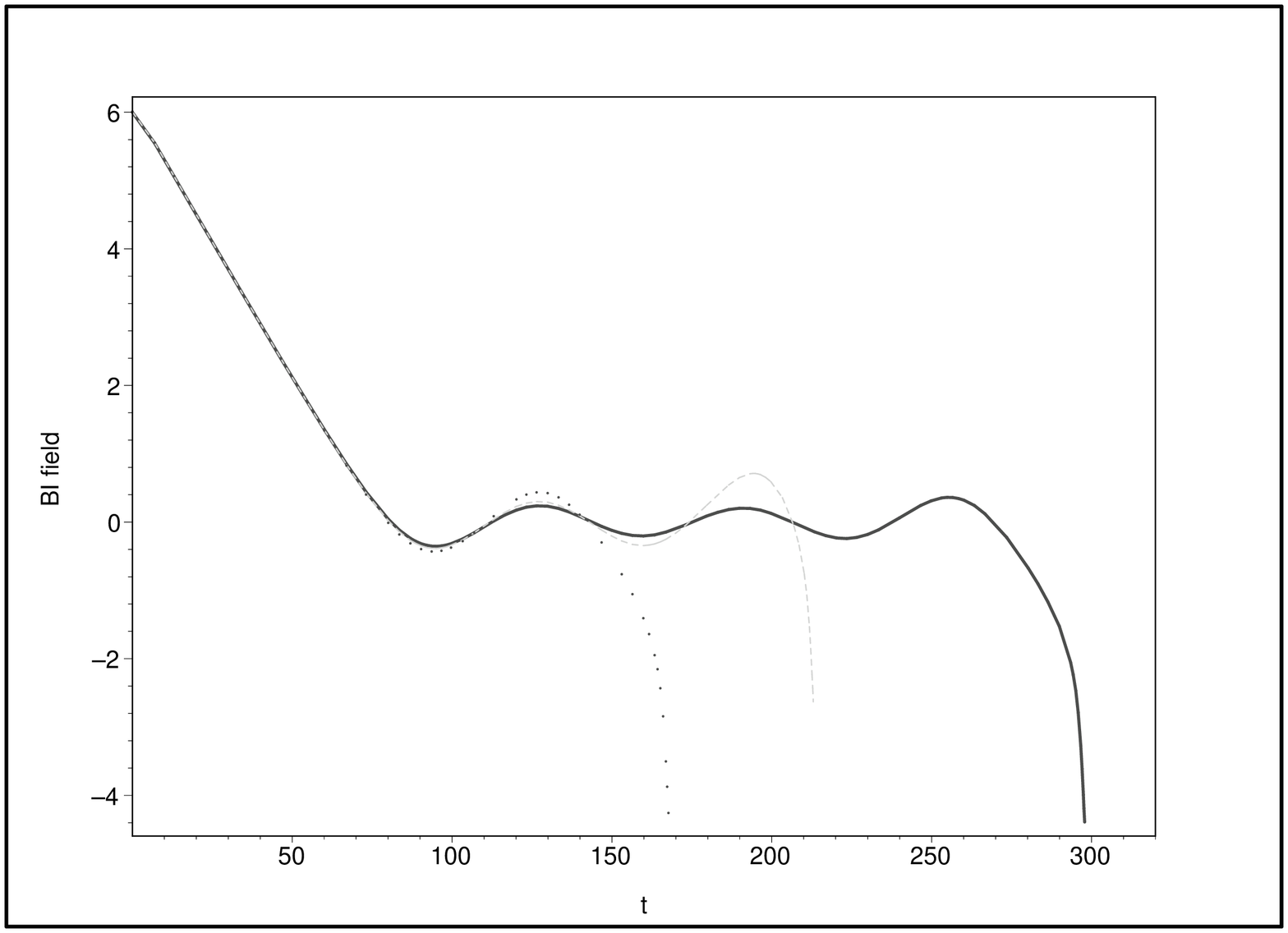}
{\small  ~Fig5.The evolution of scalar field $\phi$ with the three
potentials. The initial condition and the value of parameters are
the same as Fig4.}
\end{minipage}
\hfill
\begin{minipage}{0.47\textwidth}
\includegraphics[scale=0.32,origin=c,angle=270]{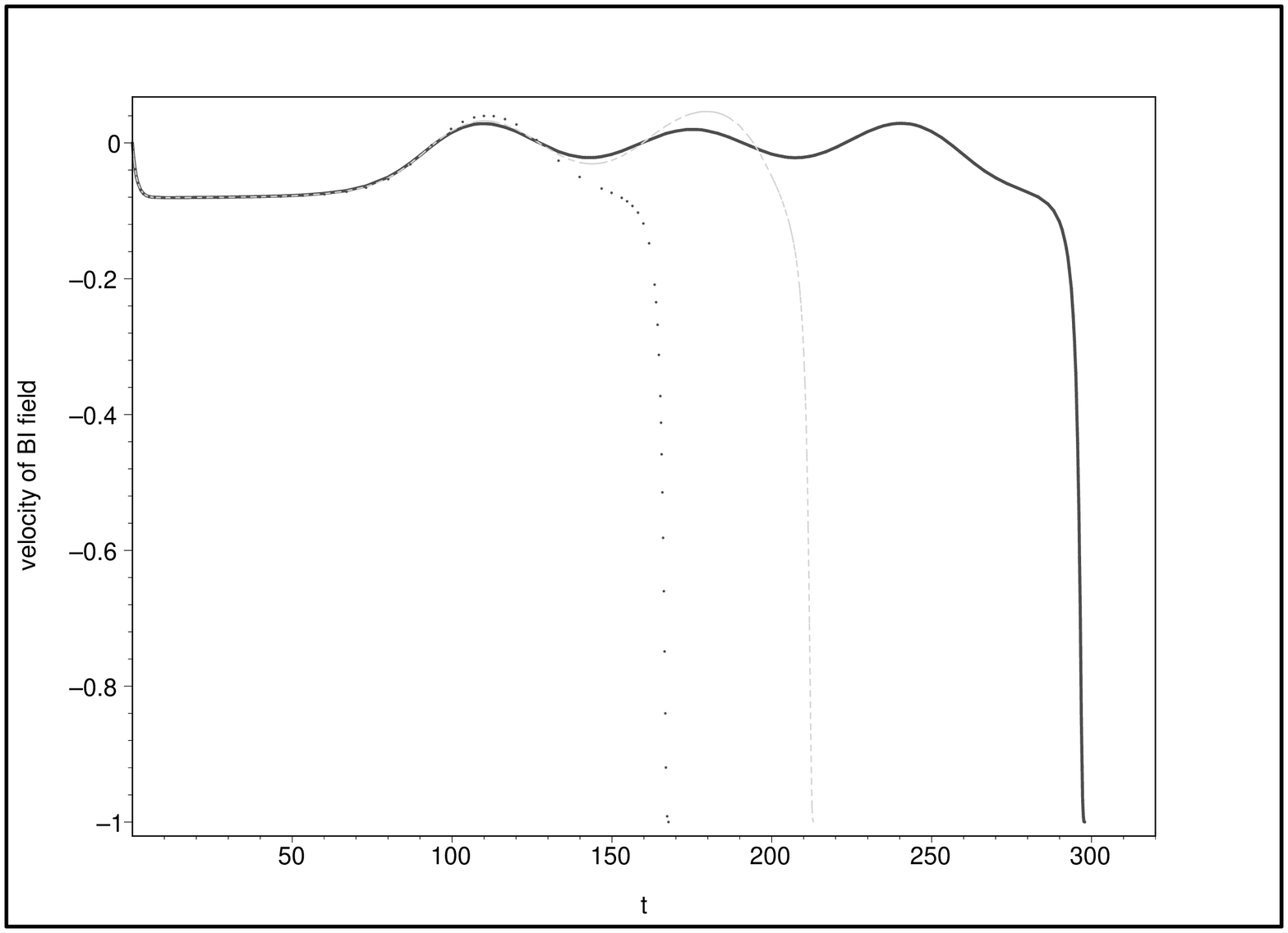}
{\small~Fig6.The evolution of field velocity $\dot\phi$ with the
three potentials. The initial condition and the value of
parameters are the same as Fig4.}
\end{minipage}
\end{center}
\par \textbf{Case 2. Same potential well deep $V_0$ but different
slope $m$  }
\begin{center}\vspace{0.5cm}
\begin{minipage}{0.48\textwidth}
\includegraphics[scale=0.32,origin=c,angle=270]{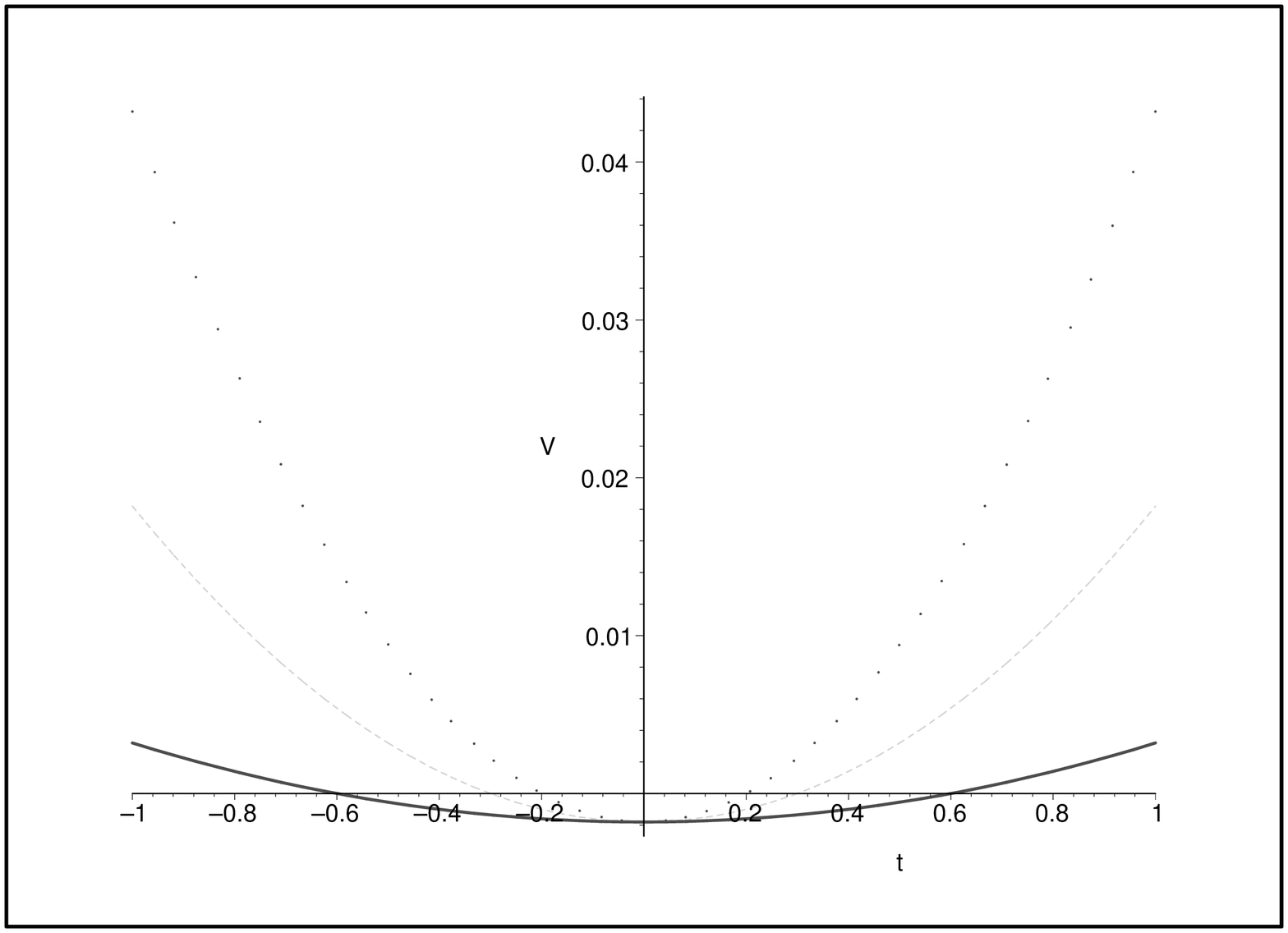}
{\small~Fig7. The three potentials with same potential well
$V_0(=-0.0018)$ but different slope $m$. $m=0.1$ for solid line,
$m=0.2$ for dash line and $m=0.3$ for dot line.}
\end{minipage}
\hfill
\begin{minipage}{0.47\textwidth}
\includegraphics[scale=0.32,origin=c,angle=270]{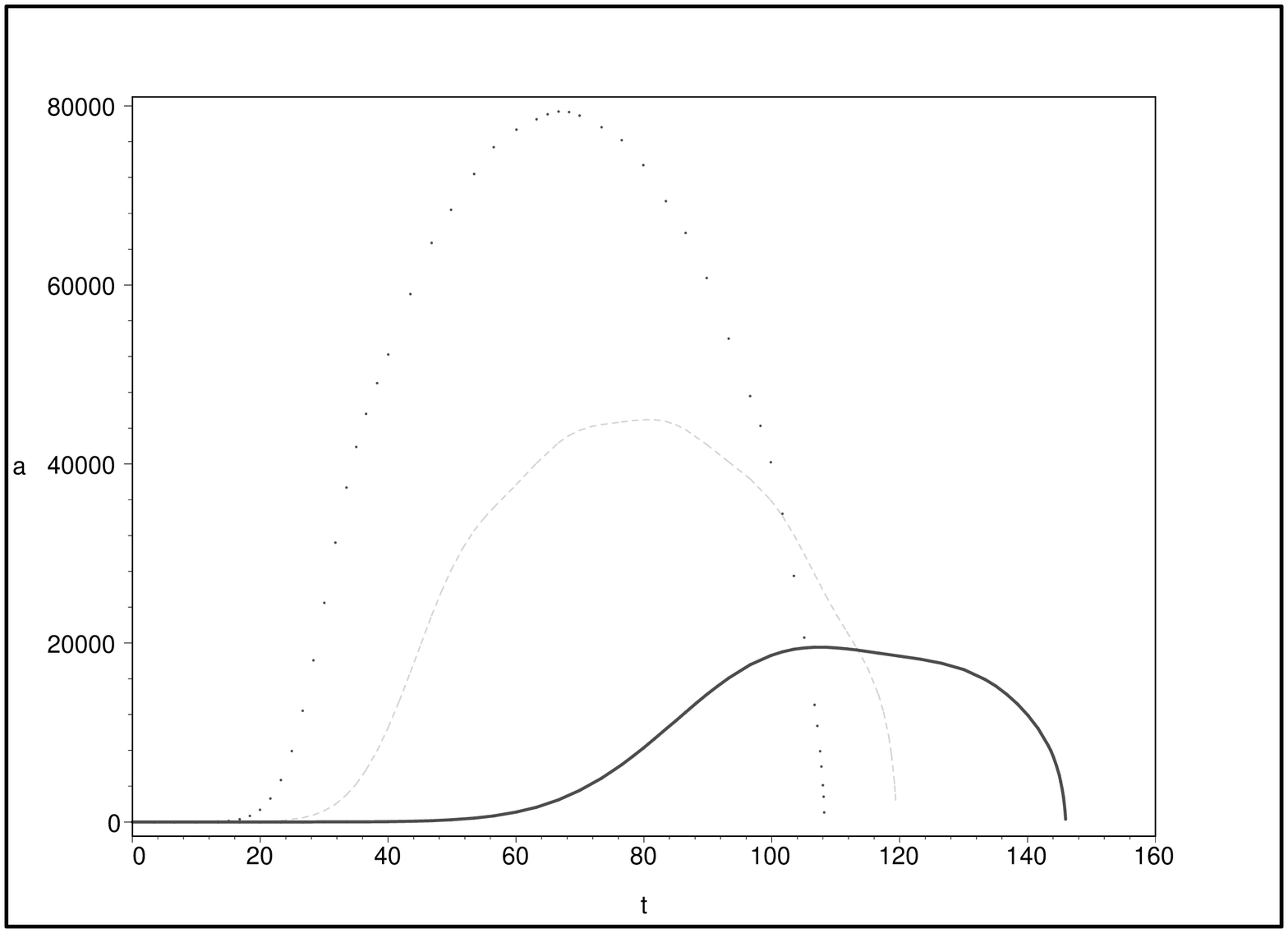}
{\small~Fig8. The evolution of scale factor $a(t)$ with the three
potentials. The value of parameters are the same as Fig.7. The
initial condition is $\phi(0)=8$, $\dot\phi(0)=0$ }
\end{minipage}
\hspace{0.4\textwidth}
\begin{minipage}{0.48\textwidth}
\includegraphics[scale=0.32,origin=c,angle=270]{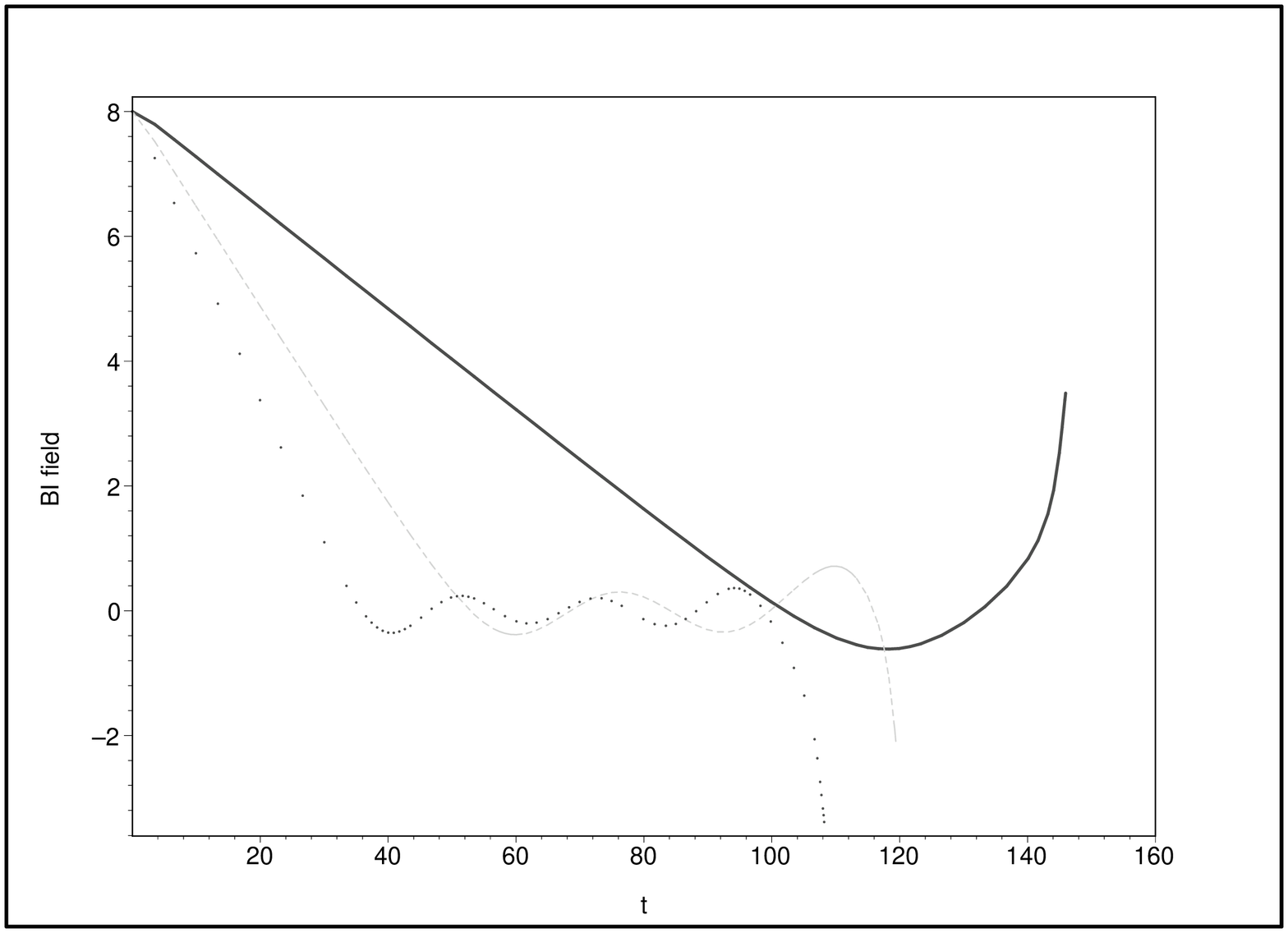}
{\small~Fig9.The evolution of scalar field $\phi$ with the three
potential. The initial condition and the value of parameters are
the same as Fig.8.}
\end{minipage}
\hfill
\begin{minipage}{0.48\textwidth}
\includegraphics[scale=0.32,origin=c,angle=270]{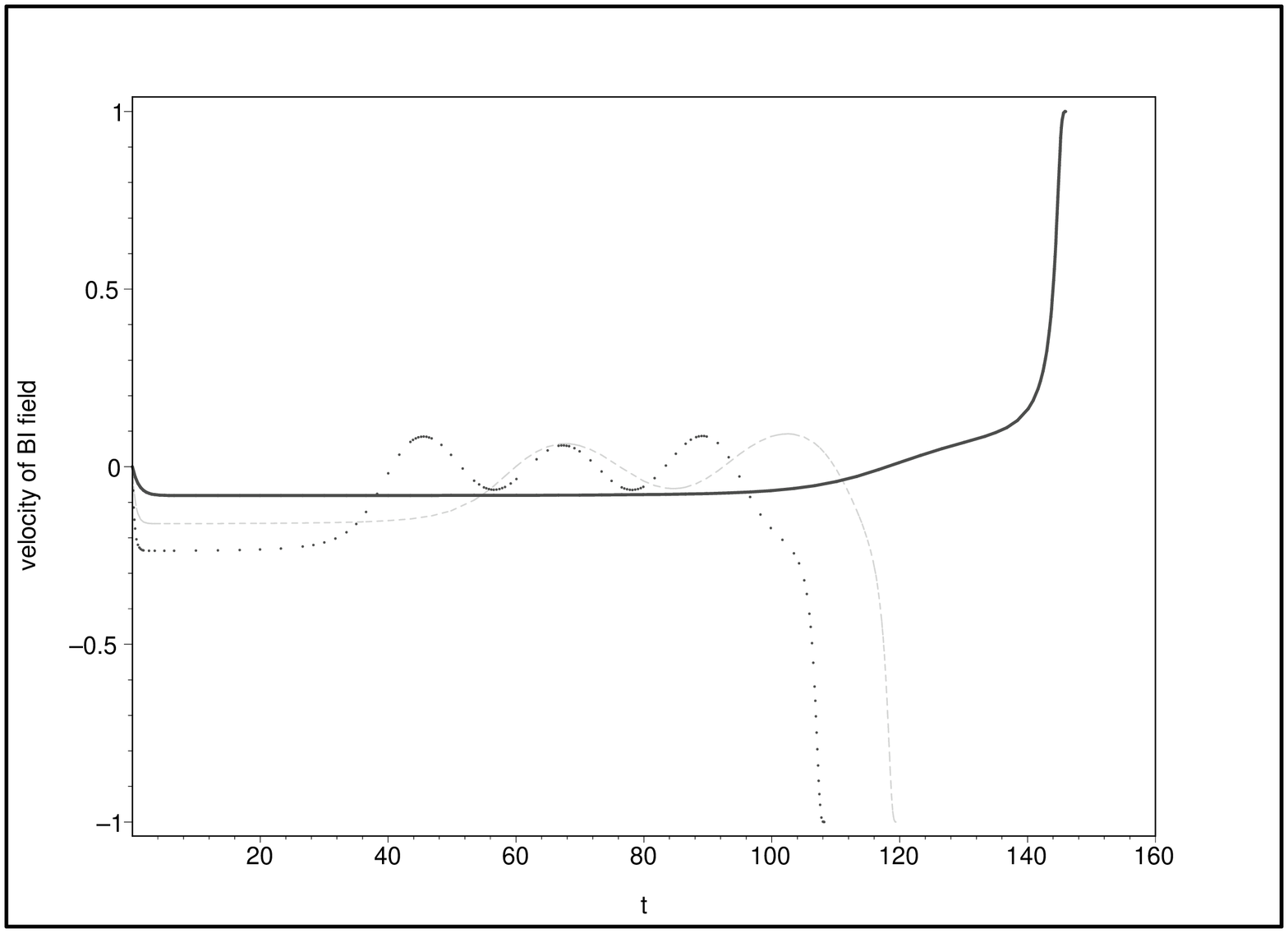}
{\small~Fig10.The evolution of field velocity $\dot\phi$ with the
three potential. The initial condition and the value of parameters
are the same as Fig.8.}
\end{minipage}
\end{center}
\par The result presented in Fig.8 shows that
the steeper the potential slope is, the shorter the universe age
is. Therefore there also exist a upper bound for the value of
slope $m$ to accommodate an accelerating expansive universe with
the large scale structure formed. At the time when universe begins
to contract from an expansion, the scalar field $\phi$ begins to
oscillate and ultimate moves to $\pm\infty$(Fig.9). The velocity
of field $|\dot\phi|$ will reach its maximum value $1$ finally.
\par \textbf{Case 3. Comparison between Linear Scalar Field and NLBI Scalar Field }
\par In NLBI scalar field
theory, the linear scalar field theory is considered only
correctly in weak field regime and will be not valid in strong
field regime. It is necessary to investigate their different
cosmological evolution in this two theories with the same
parameter value and initial conditions. The results are plotted in
Figs.11-14.
\par For the same parameter value and initial condition, Figs.11-14
show that this difference is quite noticeable. The different
evolution of scale factor(Fig.12) leads to different expansive
rate $H_i$ at time $i$(Fig.11): the value of Hubble parameter in
NLBI scalar field theory is lager than the value in linear scalar
field.
 \vskip 0.3in
\begin{center}\vspace{0.5cm}
\begin{minipage}{0.48\textwidth}
\includegraphics[scale=0.32,origin=c,angle=270]{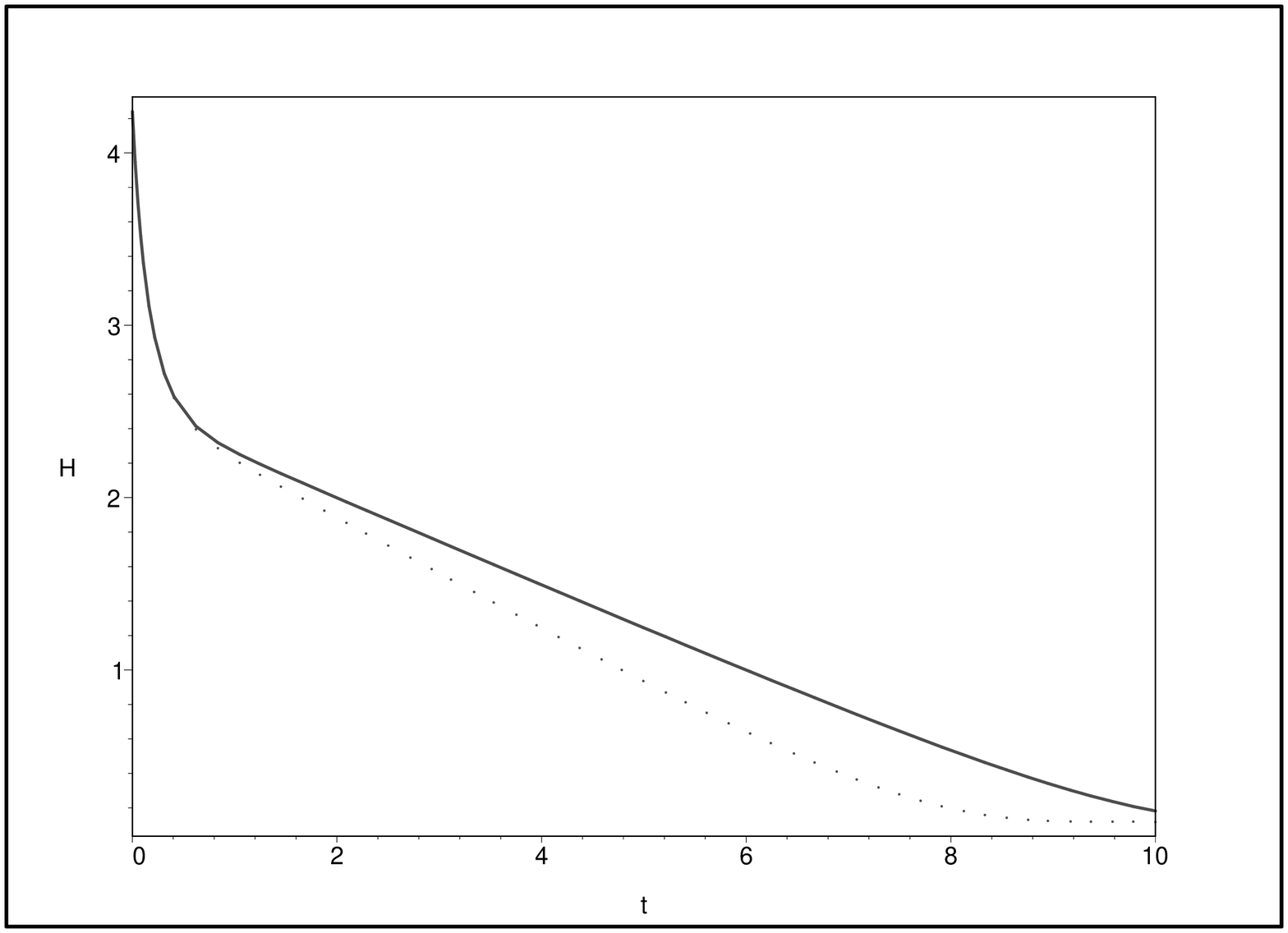}
{\small~Fig11. The evolution of Hubble parameter $H$ with respect
to $t$. Solid line for NLBI scalar field, dot line for linear
scalar field. The value of parameters
is chosen for $m=1, V_0=0.02$. The initial condition is $\phi(0)=6$, $\dot\phi(0)=0$.\\
}
\end{minipage}
\hfill
\begin{minipage}{0.48\textwidth}
\includegraphics[scale=0.32,origin=c,angle=270]{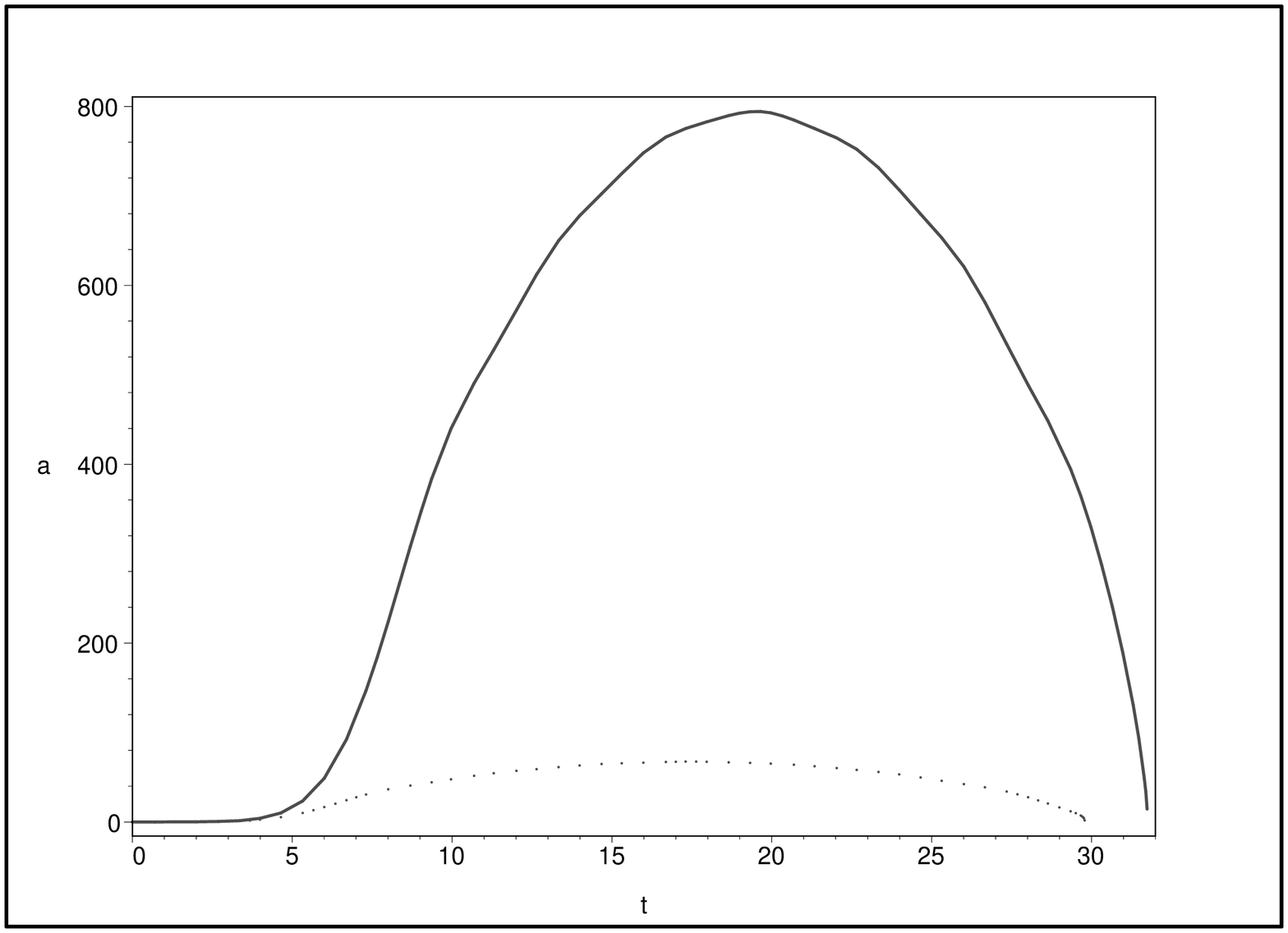}
{\small~Fig12. The evolution of scale factor $a$ with respect to
$t$. Solid line for NLBI scalar field theory, dot line for linear
scalar field theory. The value of parameters and the initial
condition is the same as Fig.11.\\ }
\end{minipage}
\hfill
\begin{minipage}{0.48\textwidth}
\includegraphics[scale=0.32,origin=c,angle=270]{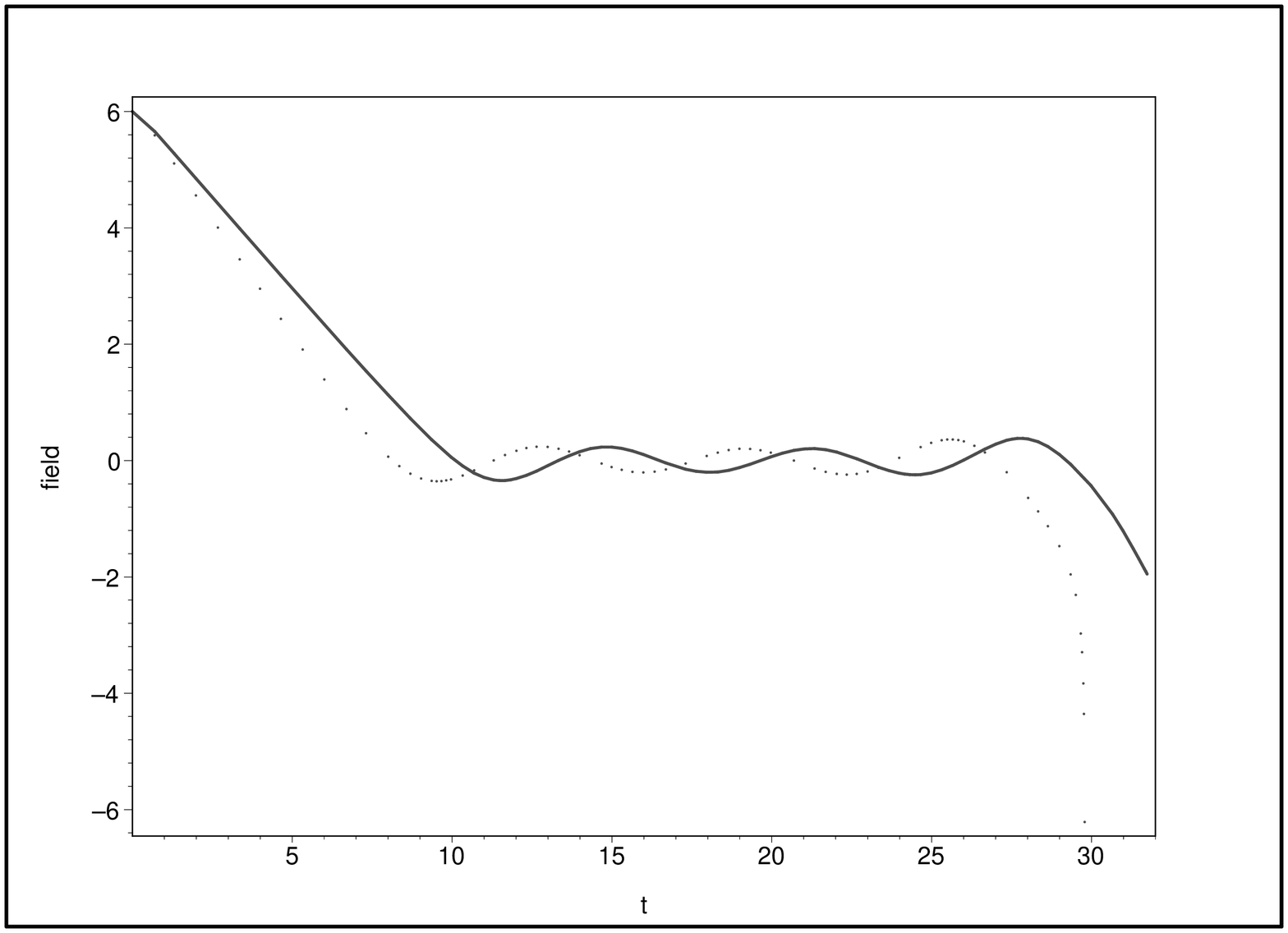}
{\small ~Fig13. The evolution of scalar field $\phi$ with respect
to $t$. Solid line for NLBI scalar field theory, dot line for
linear scalar field theory. The value of parameters and the
initial condition is the same as Fig.11.\\}
\end{minipage}
\hfill
\begin{minipage}{0.48\textwidth}
\includegraphics[scale=0.32,origin=c,angle=270]{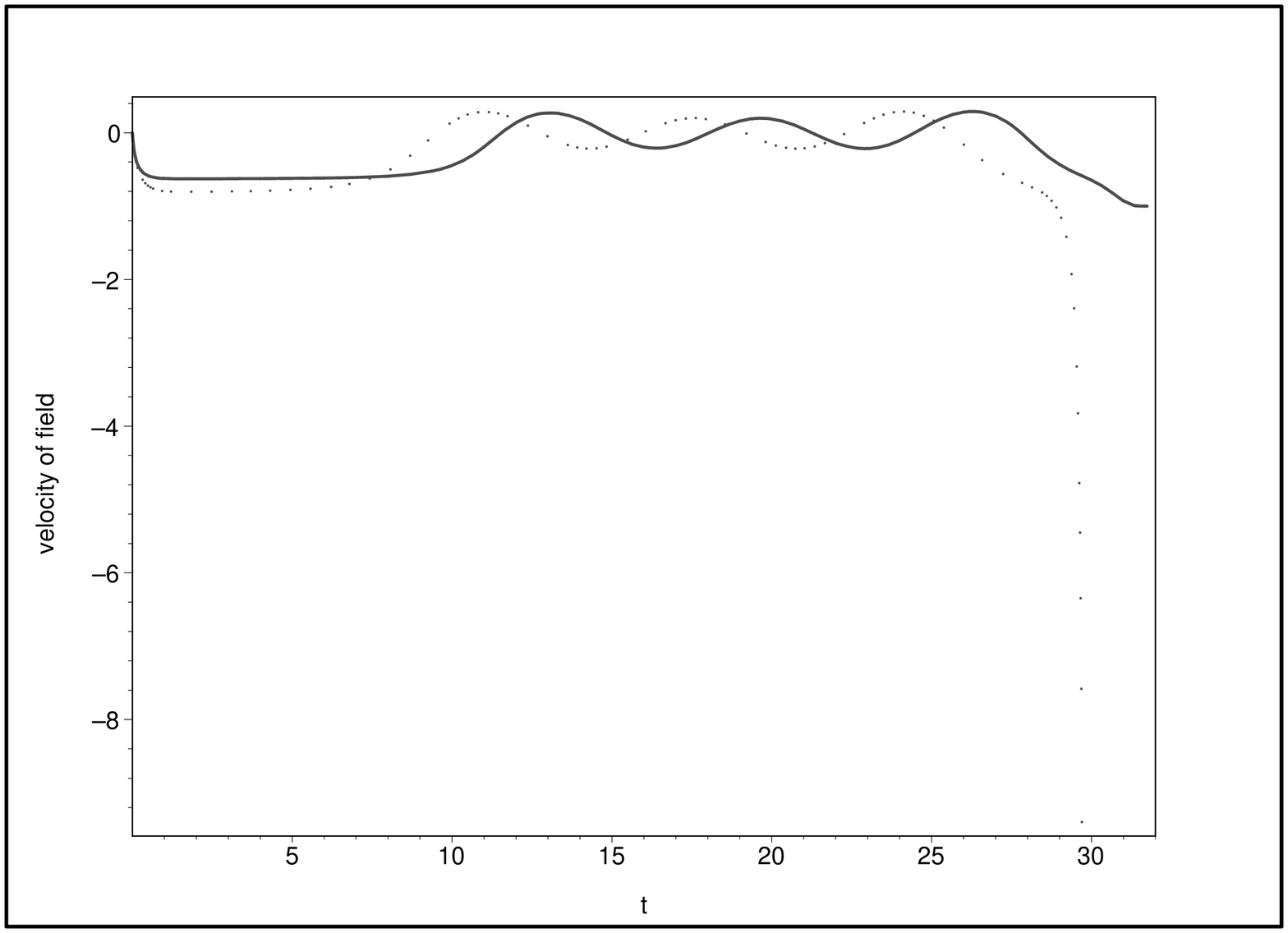}
{\small~Fig14. The evolution of scalar field $\dot\phi$ with
respect to $t$. Solid line for NLBI scalar field theory, dot line
for linear scalar field theory. The value of parameters and the
initial condition is the same as Fig.11.\\}
\end{minipage}
\end{center}

\par \textbf{Case 4. Comparison between positive, non-negative and negative potentials }
\par Now we have known that whether the potential can evolve to negative value is very important to the destiny of universe.
A spatially flat universe with a negative potential may eventually
collapses, which is not the same as in the general textbook. We
plot the different cosmological evolution with the potential
parameter $V_0>0$, $V_0=0$, $V_0<0$ in NLBI scalar field theory.
The results are plotted in Figs.15-17.
 \vskip 0.3 in
\begin{center}\vspace{0.55cm}
\begin{minipage}{0.5\textwidth}
\includegraphics[scale=0.32,origin=c,angle=270]{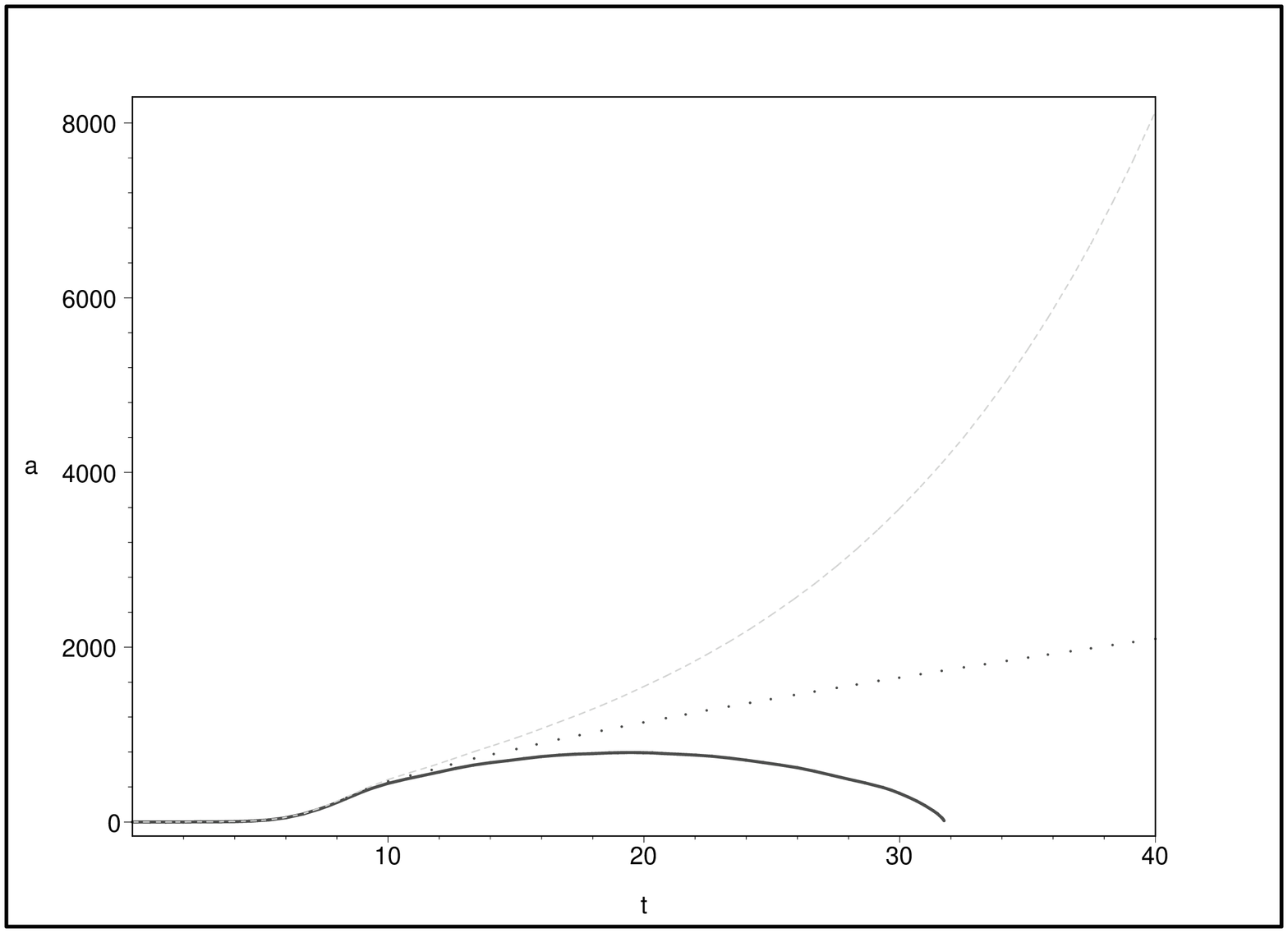}
{\small Fig15.The evolution of scale factor $a$ with respect to
$t$. Solid line for $V_0=-0.02$, dash line for $V_0=0$, dot line
for$V_0=0.02$.  $m=1$ and the initial condition is $\phi(0)=6$,
$\dot\phi(0)=0$.
\\}
\end{minipage}
\hfill
\begin{minipage}{0.4\textwidth}
\includegraphics[scale=0.32,origin=c,angle=270]{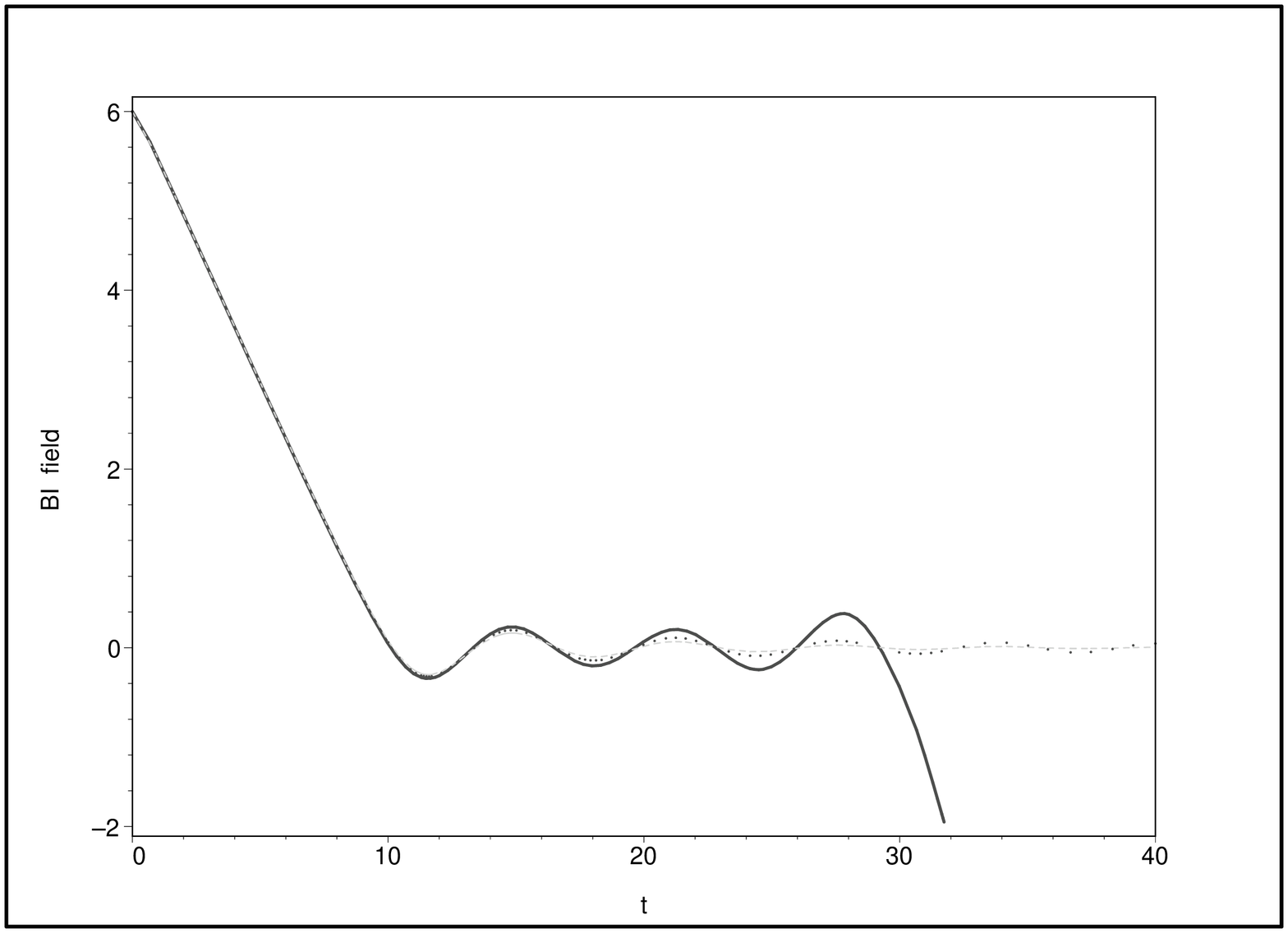}
{\small~Fig16. The evolution of scalar field $\phi$ with respect
to $t$. The value of parameter and the initial condition is the
same as Fig15.
\\}
\end{minipage}
\end{center}
\par Furthermore, we plot
the evolution of the density parameter $\Omega$ when $V_0>0,
V_0=0$ and $V_0<0$(Fig.18). The starting point is chosen at the
equipartition epoch, at which $\Omega_{Mi}=\Omega_{ri}=0.5$. We
should emphasize that in fact we plot the evolution of three cases
when $V_0=4.5\times10^{-6}$, $V_0=0$, $V_0=-4.5\times10^{-6}$ in
Fig.18, but we can not find any difference from Fig.18. It shows
that up to now there are no difference in the evolution of the
density parameter $\Omega$ for small value of $|V_0|$, though the
future of the universe is dramatically different for those three
cases(see Fig.15).
\begin{center}\vspace{0.55cm}
\begin{minipage}{0.40\textwidth}
\includegraphics[scale=0.32,origin=c,angle=270]{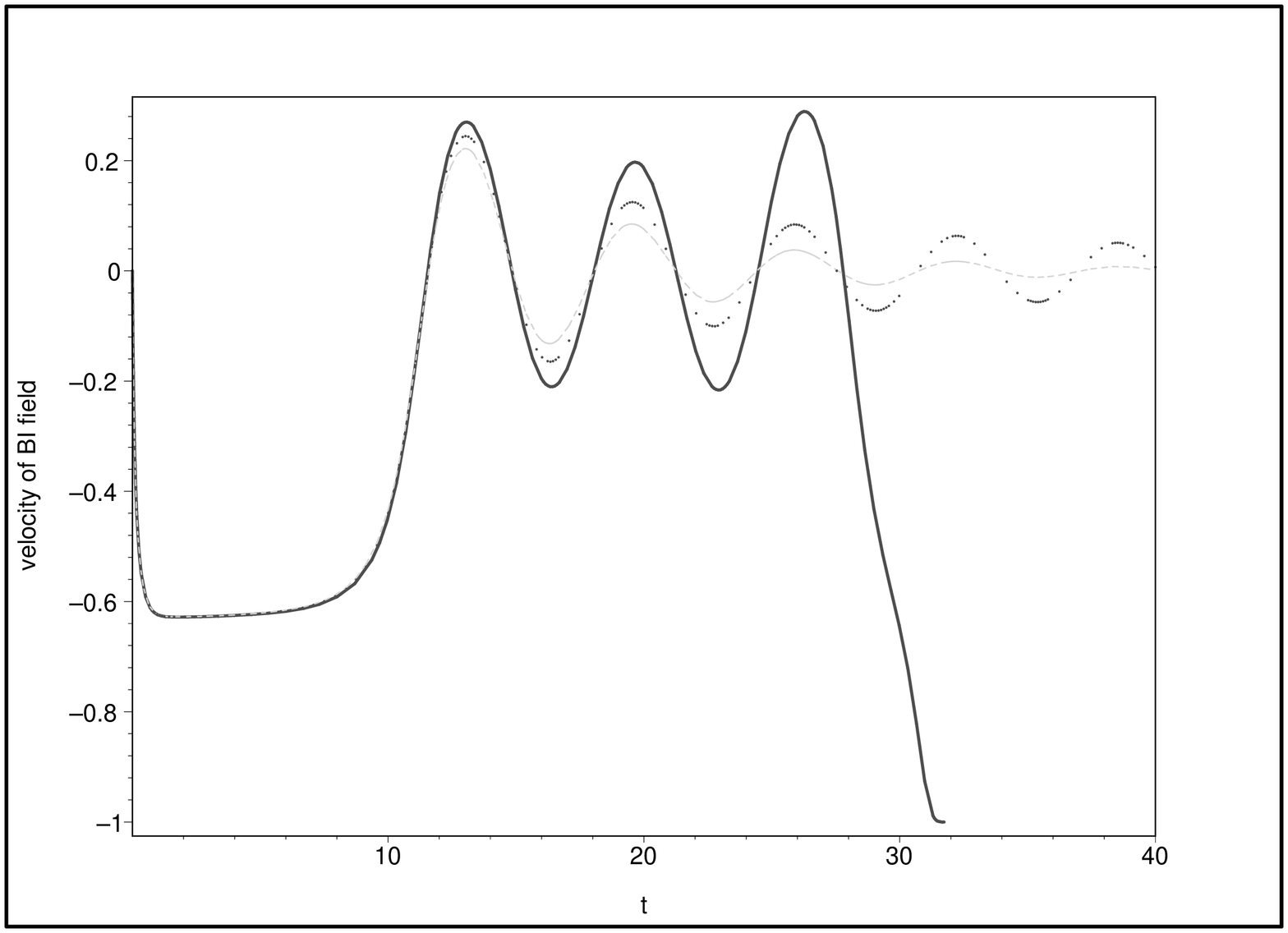}
{\small~Fig17.The evolution of field velocity $\dot\phi$ with
respect to $t$. The value of parameter and the initial condition
is the same as Fig15.\\}
\end{minipage}
\hfill
\begin{minipage}{0.50\textwidth}
\includegraphics[scale=0.32,origin=c,angle=270]{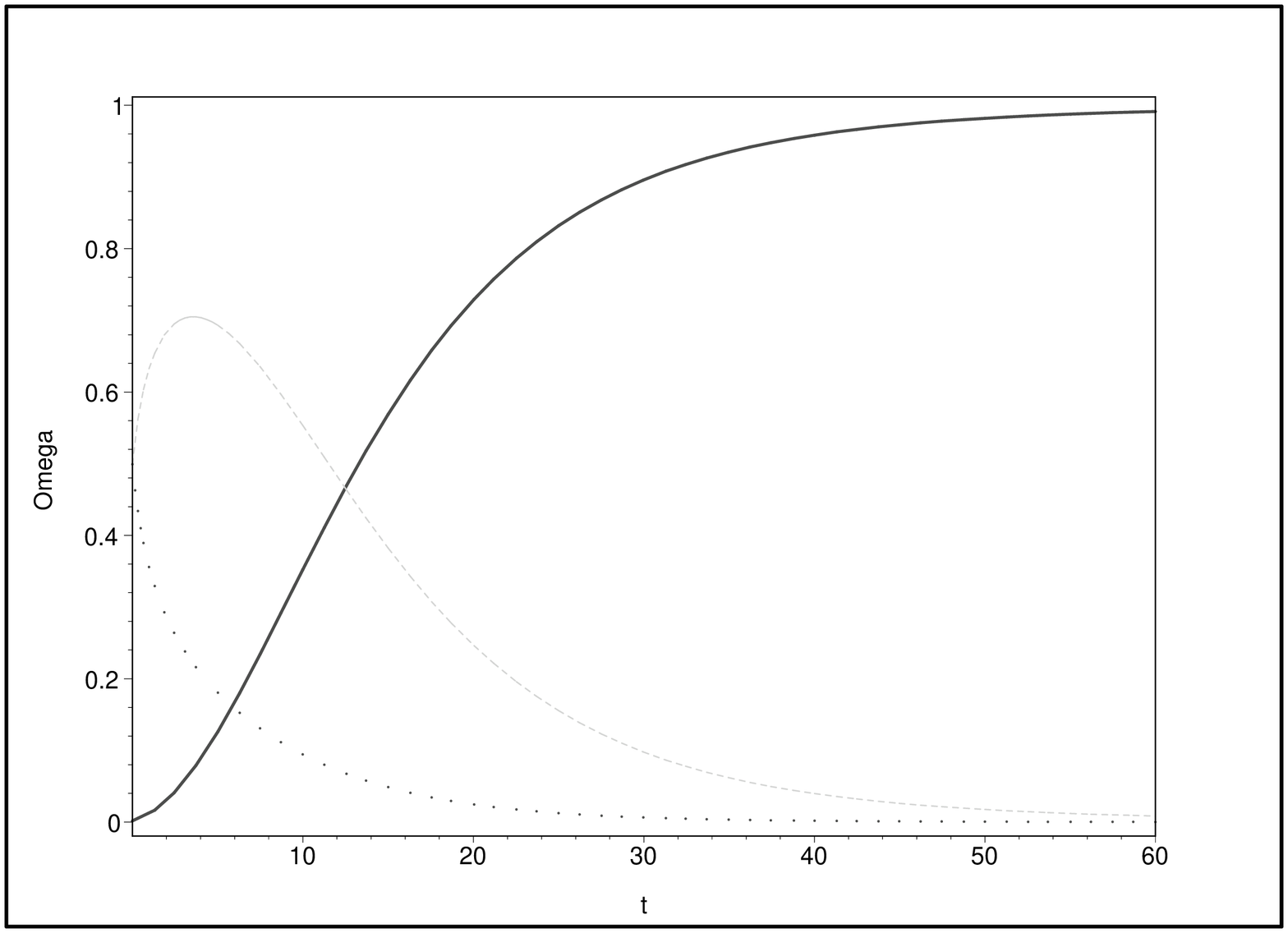}
{\small~Fig18.The evolution of cosmological density parameter
$\Omega$ with respect to $t$.  Solid line for NLBI scalar field,
dash line for matter, dot line for radiation. $m=0.03$,
$\phi(0)=6$, $\dot\phi(0)=0$
\\}
\end{minipage}
\end{center}

\section{Negative potentials and Cyclic Universe}
\par Till now we have studied the evolution of
the universe and classified new possibilities which appear in NLBI
scalar field theories with negative potentials. It is very
interesting to investigate its cosmological evolution and its
differences with linear scalar field. It is true that the universe
with negative potentials will end with a crunch and never expand
again. They are born in a singularity and end in a singularity.
\par Recently, P.J.Steinhardt and N.Turok proposed a version of
cyclic scenario[80-82]. It is based on the idea that we live on one
of two branes whose separation can be parametrized by a scalar field
$\phi$. It is assumed that one can describe the brane interaction by
an effective 4D theory with the effective potential $V(\phi)$ having
a negative minimum. According to brane-view,  the potential
$V(\phi)$ is the inter-brane potential caused non-perturbative
virtual exchange of membranes between the boundaries. The interbrane
force is what causes the branes to repeatedly collide and bounce.
Consequently, the scale factor bounces and begins to expand. It is
assumed that the potential $V(\phi)$ equals an extremely small
value($\sim10^{-120}$) at large $\phi$, and therefore the universe
experiences a stage of extremely low-scale inflation associated with
present stage of accelerating expansion. In the cyclic universe
scenario the perturbations responsible for the formation of the
structure of the universe are produced during the contracting regime
of precious cycle. After the new cycle creates from singularity, the
universe will experience radiation and matter domination, a
low-scale inflation(i.e, dark energy domination) and contract again.
Obviously there is no inflation in the very early universe. The
author claimed that the cyclic model is able to reproduce all of the
successful predictions of the consensus model (inflationary+Big Bang
cosmology) with the same exquisite detail.
 \par However, later, G.Felder, A.Frolov, L.Kofman and A.Linde[77] investigate the
cyclic universe and show that there are some problems that need to
be resolved in order to realize a cyclic regime in this scenario.
They propose several modifications of this scenario and conclude
that the best way to improve it is to add a usual stage of inflation
after the singularity and use the inflationary stage to generate
perturbations in the standard way. In fact the inflationary
mechanism is not the alternative to big bang model, it can be
accommodated into big bang model instead. So in order to solve some
problems appeared in cyclic model we can also involve the
inflationary mechanism, just as G.Felder, A.Frolov, L.Kofman and
A.Linde have suggested in Ref[77]. In this cyclic model we will find
that all the cosmological detail in the consensus model will appear.
Very recently[83], P.J.Steinhardt and N.Turok show that the cyclic
model can naturally incorporate a dynamical mechanism that
automatically relaxes the value of the cosmological constant. It can
explain why the cosmological constant is small and positive, as
observed today.
\section{Conclusion and Summary }
The main goal of this paper is to perform a general investigation
of the NLBI scalar field cosmology with negative potentials. The
cosmological solutions in different regime have obtained through
some approximate approach. The results obtained in NLBI scalar
field theory are quite different with that obtained in linear
scalar field theory. A notable characteristic is that NLBI scalar
field behaves as ordinary matter nearly the singularity while the
linear scalar field behaviors as "stiff" matter. We also find
that, due to the nonlinear effect, the oscillatory motion of
$\phi$ in the vicinity when the universe evolve to contraction
from expansion is different to linear scalar field. Moreover, the
value of Hubble parameter $H_i$ at time $i$ in NLBI scalar field
theory is large than the one in linear scalar field theory. With
the investigation of evolution with different value of $m$ and
$V_0$, we find that in order to accommodate an accelerating
expansive universe in which the large scale structure had formed,
the value of $m$ and $|V_0|$ must have a {\it upper bound}.
Finally we review the negative potentials and the new cyclic
model.
\section{Acknowledgement}
 \hspace*{15 pt}We appreciate the anonymous referee for the helpful suggestion. This work is partly supported by NNSFC under Grant No.10573012 and No.10575068 and by Shanghai
 Municipal Science and Technology Commission No.04dz05905. \\

{\noindent\Large \bf References} \small{
\begin{description}
\item {1.} {B.Ratra and P.J.Peebles, Phys.Rev.D\textbf{37}, 3406(1998).}
\item {2.} {R.R.Caldwell, R.Dave and P.J.Steinhardt, Phys.Rev.Lett.\textbf{80}, 1582(1998).}
\item {3.} {P.J.Steinhardt, L.Wang and I.Zlatev, Phys.Rev.Lett.\textbf{82}, 896(1996).}
\item {4.} {X.Z.Li, J.G.Hao and D.J.Liu, Class.Quant.Grav.\textbf{19}, 6049(2002).}
\item {5.} {X.Z.Li, J.G.Hao, D.J.Liu and X.H.Zhai, Int.J.Mod.Phys.A\textbf{18}, 5921(2003).}
\item {6.} {Y.G.Gong, Int.J.Mod.Phys.D\textbf{15}, 599(2005).}
\item {7.} {C.A.Picon, T.Damour and V.Mukhanov, Phys.Lett.B\textbf{458}, 209(1999).}
\item {8.} {C.A.Picon, V.Mukhanov and P.J.Steinhardt, Phys.Rev.Lett.\textbf{85}, 4438(2000).}
\item {9.} {J.M.Aguirregabiria, L.P.Chiemento and R.Lazkoz, Phys.Rev.D\textbf{70}, 023509(2004).}
\item {10.} {L.P.Chimento, Phys.Rev.D\textbf{69}, 123517(2004).}
\item {11.} {J.Garriga and V.F.Mukhanov, Phys.Lett.B\textbf{458}, 219(1999).}
\item {12.} {A.G.Riess, astro-ph/0402512.}
\item {13.} {G.Panotopoulos, astro-ph/0606249.}
\item {14.} {C.Armend\'{a}riz-Pic\'{o}n, V.Mukhanov and P.J.Steinhardt, Phys.Rev.Lett\textbf{85}, 4438(2000).}
\item {15.} {C.Armend\'{a}riz-Pic\'{o}n, V.Mukhanov and P.J.Steinhardt, Phys.Rev.D\textbf{63}, 103510(2001).}
\item {16.} {C.Armend\'{a}riz-Pic\'{o}n, T.Damour and V.Mukhanov, Phys.Lett.B\textbf{458}, 209(1999).}
\item {17.} {T.Chiba, Phys.Rev.D\textbf{66}, 063514.}
\item {18.} {T.Chiba, T.Okabe and M.Yamaguchi, Phys.Rev.D\textbf{62}, 023511(2000).}
\item {19.} {M.Malquarti, E.J.Copeland, A.R.Liddle and M.Trodden, Phys.Rev.D\textbf{67}, 123503(2003).}
\item {20.} {R.J.Sherrer, Phys.Rev.Lett.\textbf{93}, 011301(2004).}
\item {21.} {E.J.Copeland, M.R.Garousi, M.Sami and S.Tsujikawa, Phys.Rev.D\textbf{71}, 043003(2005).}
\item {22.} {A.Melchiorri, L.Mersini, C.J.Odman and M.Trodden, Phys.Rev.D\textbf{68}, 043509(2003).}
\item {23.} {R.R.Caldwell, Phys.Lett.B\textbf{545}, 23(2002).}
\item {24.} {M.Sami, Mod.Phys.Lett.A\textbf{19}, 1509(2004).}
\item {25.} {M.Sami and T.Padamanabhan, Phys.Rev.D\textbf{67}, 083509(2003).}
\item {26.} {M.Sami, P.Chingangbam and T.Qureshi, hep-th/0301140.}
\item {27.} {P.Singh, M.Sami and N.Dadhich, Phys.Rev.D\textbf{68}, 023522(2003).}
\item {28.} {V.Faraoni, Class.Quant.Grav.\textbf{22}, 3235(2005).}
\item {29.} {D.J.Liu and X.Z.Li, Phys.Rev.D\textbf{68}, 067301(2003).}
\item {30.} {M.P.Dabrowski et.al., Phys.Rev.D\textbf{68}, 103519(2003).}
\item {31.} {S.M.Carroll, M.Hoffman and M.Teodden, astro-th/0301273.}
\item {32.} {Y.S.Piao, R.G.Cai, X.M.Zhang and Y.Z.Zhang, hep-ph/0207143.}
\item {33.} {S.Mukohyama, Phys.Rev.D\textbf{66}, 024009(2002).}
\item {34.} {T.Padmanabhan, Phys.Rev.D\textbf{66}, 021301(2002).}
\item {35.} {G.Shiu and I.Wasserman, Phys.Lett.B\textbf{541}, 6(2002).}
\item {36.} {L.kofman and A.Linde, hep-th/020512.}
\item {37.} {A.Sen, hep-th/0207105.}
\item {38.} {I.Ya.Aref'eva and A.S.Koshelev, hep-th/0605085.}
\item {39.} {I.Ya.Aref'eva, A.S.Koshelev and S.Y.Vernov, Phys.Rev.D\textbf{72}, 064017(2005).}
\item {40.} {I.Ya.Aref'eva, A.S.Koshelev and S.Y.Vernov, Phys.Lett.B\textbf{628}, 1(2005).}
\item {41.} {I.Ya.Aref'eva, A.S.Koshelev and S.Y.Vernov, astro-ph/0412619.}
\item {42.} {N.Moeller and B.Zwiebach, JHEP\textbf{0210}, 034(2002).}
\item {43.} {P.Mukhopadhyay and A.Sen, hep-th/020814.}
\item {44.} {T.Okunda and S.Sugimoto, hep-th/0208196.}
\item {45.} {G.Gibbons, K.Hashimoto and P.Yi, hep-th/0209034.}
\item {46.} {B.Chen, M.Li and F.Lin, hep-th/0209222.}
\item {47.} {G.Felder, L.Kofman and A.Starobinsky, JHEP\textbf{0209}, 026(2002).}
\item {48.} {M.C.Bento, O.Bertolami and A.A.Sen, hep-th/020812.}
\item {49.} {H.Lee et.al., hep-th/0210221.}
\item {50.} {J.G.Hao and X.Z.Li, Phys.Rev.D\textbf{66}, 087301(2002).}
\item {51.} {J.G.Hao and X.Z.Li, Phys.Rev.D\textbf{68}, 043501(2003).}
\item {52.} {X.Z.Li and X.H.Zhai, Phys.Rev.D\textbf{67}, 067501(2003).}
\item {53.} {A.de la Macorra and C.Stephan, Phys.Rev.D\textbf{65}, 083520(2002).}
\item {54.} {Imogen P.C.Heard and David Wands, Class.Quant.Grav.\textbf{19}, 5435(2002).}
\item {55.} {A.de la Macorra and G.German, Int.J.Mod.Phys.D\textbf{13}, 1939(2004).}
\item {56.} {L.Perivolaropoulos, Phys.Rev.D\textbf{71}, 063503(2005).}
\item {57.} {W.Heisenberg, Z.Phys.\textbf{133},79(1952).}
\item {58.} {W.Heisenberg, Z.Phys.\textbf{126}, 519(1949).}
\item {59.} {W.Heisenberg, Z.Phys.\textbf{113}, 61(1939).}
\item {60.} {G.W.Gibbons and C.A.R.Herdeiro, Phys.Rev.D\textbf{63}, 064006(2001).}
\item {61.} {G.W.Gibbons, Rev.Mex.Fis.\textbf{49}S1, 19-29(2003)(hep-th/0106059).}
\item {62.} {V.V.Dyadichev, D.V.Gal'tsov and A.G.Zorin, Phys.Rev.D\textbf{65}, 084007(2002).}
\item {63.} {D.N.Vollick, Gen.Rel.Grav.\textbf{35}, 1511(2003).}
\item {64.} {D.N.Vollick, Phys.Rev.D\textbf{72}, 084026(2005).}
\item {65.} {H.P.de Oliveira, J.Math.Phys.\textbf{36}, 2988(1995).}
\item {66.} {T.Taniuti, Prog.Theor.Phys.(kyoto) Suppl\textbf{9}, 69(1958).}
\item {67.} {H.Q.Lu, T.Harko and K.S.Cheng, Int.J.Mod.Phys.D\textbf{8}, 625(1999).}
\item {68.} {H.Q.Lu et al., Int.J.Theor.Phys\textbf{42}, 837(2003).}
\item {69.} {H.Q.Lu, Int.J.Mod.Phys.D\textbf{14}, 355(2005).}
\item {70.} {W.Fang, H.Q.Lu, Z.G.Huang and K.F.Zhang, Int.J.Mod.Phys.D\textbf{15}, 199(2006).}
\item {71.} {W.Fang, H.Q.Lu, B.Li and K.F.Zhang, Int.J.Mod.Phys.D\textbf{15}, 1947(2006).}
\item {72.} {V.Mukhanov and A.Vikman, JCAP\textbf{0602}004 (2006).}
\item {73.} {A.Vikman, astro-ph/0606033.}
\item {74.} {E.Babichev, V.Mukhanov and A.Vikman, hep-th/0604075}
\item {75.} {A.F$\ddot u$zfa and J.M.Alimi, Phys.Rev.D\textbf{73}, 023520(2006).}
\item {76.} {A.F$\ddot u$zfa and J.M.Alimi, Phys.Rev.Lett.\textbf{97}, 061301(2006).}
\item {77.} {G.Felder, A.Frolov, L.Kofman and A.Linde, Phys.Rev.D\textbf{42}, 023507(2002).}
\item {78.} {A.D.Linde, Phys.Lett.B\textbf{129}, 177(1983).}
\item {79.} { A.D.Linde, $Particle$ $Physics$ $and$ $Inflationary$ $Cosmology$(Harwood), Chur, Switzerland, 1990.}
\item {80.} {P.J.Steinhart and N.Turok, hep-th/0111030.}
\item {81.} { P.J.Steinhart and N.Turok, astro-ph/0204479.}
\item {82.} { P.J.Steinhardt and N.Turok, Phys.Rev.D\textbf{65}, 126003(2002).}
\item {83.} {P.J.Steinhart and N.Turok, astro-ph/0605173.}
\end{description}}
\end{document}